\documentclass[twocolumn]{aastex62}

\NoNewPageAfterKeywords

\usepackage[utf8]{inputenc}
\DeclareUnicodeCharacter{2212}{-}
\usepackage{graphicx}
\graphicspath{{./figures/}{./}}
\usepackage{amsmath}
\usepackage{afterpage}
\usepackage{enumitem}
\setenumerate{itemsep=0mm}
\hypersetup{    
  urlcolor        = {blue},
}
\usepackage{comment}
\usepackage{multirow}

\usepackage{lineno}

\usepackage{soul} 
\usepackage{amsmath}
\usepackage{amssymb}
\usepackage{xspace}
\usepackage{xifthen}
\usepackage{eso-pic}

\definecolor{forestgreen}{HTML}{228B22}
\definecolor{urlblue}{HTML}{000000}

\newcommand{\eg}{e.g.\xspace}

\newcommand{\Gaia}{{\it Gaia}\xspace}
\newcommand{\SSSSS}{${S}^5$\xspace}

\mathchardef\mhyphen="2D

\newcommand{\roughly}{\ensuremath{ {\sim}\,} }

\newlength{\dhatheight}

\newcommand{\code}[1]{\texttt{#1}\xspace}

\newcommand{\unit}[1]{\ensuremath{\mathrm{\,#1}}\xspace}

\newcommand{\mas}{\unit{mas}}

\newcommand{\km}{\unit{km}}
\newcommand{\kms}{\km \second^{-1}}

\newcommand{\kpc}{\unit{kpc}}
\newcommand{\second}{\unit{s}}

\newcommand{\Msun}{\unit{M_\odot}}

\newcommand{\secref}[1]{Section~\ref{sec:#1}}
\newcommand{\appref}[1]{Appendix~\ref{app:#1}}
\newcommand{\tabref}[1]{Table~\ref{tab:#1}}

\newcommand{\figref}[1]{Figure~\ref{fig:#1}}

\newcommand{\bandvar}[2][]{%
  \ifthenelse{\isempty{#1}}{\var{#2}}{\var{#2\_#1}}%
}

\newcommand{\SExtractor}{\code{SExtractor}}

\newcommand{\var}[1]{\ensuremath{\texttt{\MakeUppercase{#1}}}\xspace}

\providecommand\physrep{\ref@jnl{Phys.~Rep.}}%
\providecommand\apjs{\ref@jnl{ApJS}}%
\providecommand{\jcap}{\ref@jnl{JCAP}}%

\begin{document}

\title{Proper Motions of Stellar Streams Discovered in the Dark Energy Survey}


\author{N.~Shipp}
\altaffiliation{LSSTC Data Science Fellow}
\affiliation{Department of Astronomy and Astrophysics, University of Chicago, Chicago IL 60637, USA}
\affiliation{Kavli Institute for Cosmological Physics, University of Chicago, Chicago, IL 60637, USA}
\affiliation{Fermi National Accelerator Laboratory, PO Box 500, Batavia, IL 60510, USA}

\author{T.~S.~Li}
\altaffiliation{Hubble Fellow}
\affiliation{Fermi National Accelerator Laboratory, PO Box 500, Batavia, IL 60510, USA}
\affiliation{Kavli Institute for Cosmological Physics, University of Chicago, Chicago, IL 60637, USA}
\affiliation{Observatories of the Carnegie Institution for Science, 813 Santa Barbara St., Pasadena, CA 91101, USA}
\affiliation{Department of Astrophysical Sciences, Princeton University, Princeton, NJ 08544, USA}

\author{A.~B.~Pace}
\affiliation{George P. and Cynthia Woods Mitchell Institute for Fundamental Physics and Astronomy, and Department of Physics and Astronomy, Texas A\&M University, College Station, TX 77843, USA}

\author{D.~Erkal}
\affiliation{Department of Physics, 12 BC 03, University of Surrey, GU2 7XH, UK}

\author{A.~Drlica-Wagner}
\affiliation{Fermi National Accelerator Laboratory, PO Box 500, Batavia, IL 60510, USA}
\affiliation{Department of Astronomy and Astrophysics, University of Chicago, Chicago IL 60637, USA}
\affiliation{Kavli Institute for Cosmological Physics, University of Chicago, Chicago, IL 60637, USA}

\author{B.~Yanny}
\affiliation{Fermi National Accelerator Laboratory, PO Box 500, Batavia, IL 60510, USA}

\author{V.~Belokurov}
\affiliation{Institute of Astronomy, University of Cambridge, Madingley Road, Cambridge CB3 0HA, UK}

\author{W.~Wester}
\affiliation{Fermi National Accelerator Laboratory, PO Box 500, Batavia, IL 60510, USA}

\author{S.~E.~Koposov}
\affiliation{McWilliams Center for Cosmology, Carnegie Mellon University, 5000 Forbes Ave, Pittsburgh, PA 15213, USA}
\affiliation{Institute of Astronomy, University of Cambridge, Madingley Road, Cambridge CB3 0HA, UK}

\author{G.~F.~Lewis}
\affiliation{Sydney Institute for Astronomy, School of Physics, A28, The University of Sydney, NSW, 2006, Australia}

\author{J.~D.~Simpson}
\affiliation{School of Physics, UNSW, Sydney, NSW 2052, Australia}

\author{Z.~Wan}
\affiliation{Sydney Institute for Astronomy, School of Physics, A28, The University of Sydney, NSW, 2006, Australia}

\author{D.~B.~Zucker}
\affiliation{Department of Physics \& Astronomy, Macquarie University, Sydney, NSW 2109, Australia\\}
\affiliation{Macquarie University Research Centre for Astronomy, Astrophysics \& Astrophotonics, Sydney, NSW 2109, Australia\\}

\author{S.~L.~Martell}
\affiliation{School of Physics, UNSW, Sydney, NSW 2052, Australia}
\affiliation{Centre of Excellence for All-Sky Astrophysics in Three Dimensions (ASTRO 3D), Australia}

\author{M.~Y.~Wang}
\affiliation{Department of Physics, Carnegie Mellon University, Pittsburgh, PA 15312, USA}


\collaboration{(\SSSSS Collaboration)}
\email{norashipp@uchicago.edu}

\begin{abstract}
We cross-match high-precision astrometric data from \Gaia DR2 with accurate multi-band photometry from the Dark Energy Survey (DES) DR1 to confidently measure proper motions for nine stellar streams in the DES footprint: Aliqa Uma, ATLAS, Chenab, Elqui, Indus, Jhelum, Phoenix, Tucana  III, and Turranburra.
We determine low-confidence proper motion measurements for four additional stellar streams: Ravi, Wambelong, Willka Yaku, and Turbio.
We find evidence for a misalignment between stream tracks and the systemic proper motion of streams that may suggest a systematic gravitational influence from the Large Magellanic Cloud. 
These proper motions, when combined with radial velocity measurements, will allow for detailed orbit modeling which can be used to constrain properties of the LMC and its affect on nearby streams, as well as global properties of the Milky Way's gravitational potential.

\end{abstract}


\section{Introduction}
\label{sec:intro}

Stellar streams, the tidal remnants of accreted globular clusters and dwarf galaxies, are powerful tools for studying the distribution of matter and the accretion history of our Galaxy \citep{Johnston:1998,Bullock:2005}. 
Stellar streams arise naturally in hierarchical models of galaxy formation; however, their low surface brightness makes them historically difficult to detect.
The advent of large sky surveys has rapidly increased the number of known streams around the Milky Way \citep[e.g.,][ and references therein]{Mateu:2018} and other nearby galaxies \citep[e.g.,][]{Zucker:2004,Martinez-Delgado:2010}.
This explosion in the known population of stellar streams promises to enable detailed statistical modeling of the stream population in the near future \citep[e.g.,][]{Bonaca:2018}.

Stellar streams are excellent dynamical tools for measuring the properties of the Milky Way, including the total enclosed mass within their orbits \citep[e.g.,][]{Gibbons:2014,Bowden:2015,Bovy:2016,Bonaca:2018} and the shape and radial profile of the gravitational field \citep{Law:2010,Erkal:2016a}. 
An individual stream can probe the Milky Way’s potential across tens of kiloparsecs \citep{Law:2010,Koposov:2010}, while a population of a dozen stellar streams with full kinematic information is expected to constrain the gravitational potential of the Milky Way at sub-percent levels \citep{Bonaca:2018}. 

Stellar streams are also sensitive tracers of perturbations in the Milky Way's gravitational field.
Streams can be used to detect perturbations in the gravitational field of the halo from known satellites \citep[e.g.][]{Vera-Ciro:2013,Gomez:2015,Erkal:2018a,Erkal:2018b} and smaller dark matter substructure \citep[e.g.][]{Ibata:2002,Johnston:2002,Yoon:2010,Carlberg:2009,Carlberg:2012, Erkal:2015}.
The Milky Way's largest satellite, the Large Magellanic Cloud (LMC), resides in a dark matter halo that may be 10\% as massive as that of the Milky Way \citep{Busha:2011, Boylan-Kolchin:2012}.
Direct measurements of the LMC mass exist only within $\roughly 9\kpc$ yielding values of $\roughly 2 \times 10^{10} \Msun$ \citep[e.g.,][]{Schommer:1992,vanderMarel:2014}; however, cosmological arguments predict that the mass of the LMC is nearly an order of magnitude larger \citep{Busha:2011, Boylan-Kolchin:2012}.
Such a large gravitational perturber located at a distance of only $50\kpc$ would have an appreciable affect on measurements of the gravitational field in the halo of the Milky Way.
Stellar streams, particularly those in spatial proximity to the LMC, offer a sensitive mechanism to independently measure the mass of the LMC at large radii \citep[e.g.,][]{Erkal:2018a,Erkal:2018b}. 

Large-area, ground-based, digital photometric surveys like the Sloan Digital Sky Survey \citep[SDSS;][]{York:2000}, Pan-STARRS \citep{Chambers:2016}, VST ATLAS \citep{Shanks:2015}, and the Dark Energy Survey \citep[DES;][]{DES:2016} have rapidly increased the number of known stellar streams \citep[e.g.,][]{Odenkirchen:2001,Grillmair:2006, Grillmair:2006b, Belokurov:2006, Grillmair:2009, Bonaca:2012, Koposov:2014, Grillmair:2014, Drlica-Wagner:2015, Balbinot:2016, Bernard:2016, Grillmair:2017, Grillmair:2017b, Myeong:2017, Shipp:2018, Jethwa:2018}. 
The population of stellar streams discovered in DES is of particular interest for constraining the gravitational field in the Milky Way's outer halo \citep{Shipp:2018}. 
The DES streams constitute some of the faintest and most distant streams discovered around the Milky Way and, due to the excellent photometry provided by DES, they can be distinguished from foreground stellar populations with unprecedented accuracy. 
Furthermore, the location of these streams in the Southern Hemisphere makes them sensitive probes of the joint potential of the Milky Way and LMC.

While deep photometric surveys are excellent for detecting faint stellar structures at large distances, additional phase space information is necessary for comprehensive dynamical modeling \citep[e.g.][]{Eyre:2009, Bowden:2015,Law:2010,Bovy:2014,Bovy:2016,Erkal:2018b}.
The 3D kinematics of faint stream stars can be obtained via a combination of proper motion measurements from high-precision astrometric surveys and radial velocity measurements from deep spectroscopic observations. 
The second data release from the \Gaia satellite \citep[\Gaia DR2;][]{Gaia:2018} provides unprecedented proper motion measurements for more that 1 billion stars.
\Gaia DR2 has enabled proper motion measurements for stellar streams at distances of tens of kiloparsecs \citep[\eg,][]{Price-Whelan:2018, Koposov:2019, Fardal:2019}, as well as joint photometric and astrometric discovery of previously unknown streams \citep{Malhan:2018a, Malhan:2018b}.

In addition to providing kinematic information, the systemic proper motions of stellar streams can also greatly improve the efficiency of target selection for spectroscopic follow-up surveys.
Proper motions can be used to discriminate likely stream members from foreground Milky Way stars and other halo stars. 
For example, the Southern Stellar Stream Spectroscopic Survey (\SSSSS; Li et al., submitted), an on-going program to map the line-of-sight velocities and metallicities of the DES streams using the 3.9-m Anglo-Australian Telescope's 2-degree-Field (2dF) fibre position and AAOmega spectrograph, efficiently selects targets following the techniques described here.

In this paper, we cross-matched data from DES DR1 and \Gaia DR2 to measure proper motions for stellar streams in the DES footprint.\footnote{We excluded the Palca stream from this analysis due to its large width and extent on the sky, which make it difficult to characterize.} 
We performed two distinct analyses that each combined precise photometry from DES DR1 with precise astrometry from \Gaia DR2.
First, we performed a simple ``by-eye'' analysis to visibly highlight the proper motion signal of stellar streams by removing the majority of the Milky Way foreground contamination with physically motivated cuts. 
Next, we performed a more statistically rigorous Gaussian mixture model (GMM) fit, in which we applied a less-strict data selection, and then fit a two-component Gaussian model in proper motion space to account for both the Milky Way foreground and the stream signal.

We detected and confirmed proper motion signatures for nine of the 14 streams (Aliqa Uma, ATLAS, Chenab, Elqui, Indus, Jhelum, Phoenix, Tucana III, Turranburra), including the most distant DES stream, Elqui, at $\roughly 50 \kpc$. 
The proper motions of eight of these streams were confirmed in preliminary data from \SSSSS (Li et al., submitted), while one of these streams (Turranburra) has a corresponding signal in the proper motion of coincident RR Lyrae stars.
We report low-confidence proper motion signatures of four additional streams (Ravi, Wambelong, Willka Yaku, and Turbio), and no significant proper motion signature for the Molonglo stream \citep{Grillmair:2017}.
Interestingly, we found that several of the DES streams have systemic proper motions that are misaligned with their tracks on the sky.
Such an offset is expected due the the perturbative gravitational influence of the LMC \citep{Erkal:2018b,Koposov:2019}.

This paper is organized as follows.
In \secref{data}, we describe our cross-matched sample of data from  DES DR1 and \Gaia DR2. In \secref{methods}, we discuss the two methods used to obtain proper motion measurements. In \secref{results}, we present our results, and in \secref{discussion}, we discuss some of the implications of our measurements.
We conclude in \secref{conclusion}.

\section{Data Preparation}
\label{sec:data}

Our data set consists of wide-area ground-based photometry from DES DR1 \citep{DES:2018} and precision space-based astrometric measurements from \Gaia DR2 \citep{Gaia:2018}. 
We performed an angular cross-match between these catalogs based on a matching radius of $0\farcs5$.
There is a systematic astrometric offset of $\roughly 150\mas$ between DES DR1 and \Gaia DR2.\footnote{The offset between DES DR1 and \Gaia DR2 is due to the fact that the DES DR1 absolute astrometry was tied to 2MASS \citep{DES:2018}.} Before performing the cross-match, we corrected the DES astrometry by fitting two 2D polynomials to the offsets in right ascension and declination between DES and \Gaia as a function of location in the DES footprint. After applying this correction we find the median offset between DES and \Gaia to be $\roughly 55$ mas.
The \Gaia DR2 source catalog consists predominantly of stellar objects.
To further ensure our population is not contaminated by galaxies, we cut on the DES quantity, \code{EXTENDED\_COADD} $= 0$, which selects high-confidence stars by comparing to the DES point spread function \citep[PSF; Section 4.5 of][]{DES:2018}. 
We found that this cut removes very few objects from our final catalog, and a looser selection on the DES star/galaxy separation (i.e., \code{EXTENDED\_COADD} $\leq 1$) had no effect on the results of this analysis.

We also removed objects with bad astrometric fits in \Gaia DR2. As described in \citet{Lindegren:2018}, we define $u \equiv ( \code{astrom\_chi2\_al} \allowbreak/( \code{astrom\_n\_good\_obs\_al} - 5))^{1/2}$, and we remove stars with $u > 1.2 \times  \allowbreak \max (1,  \allowbreak \exp( −0.2 (G − 19.5)))$. Here $\allowbreak \code{astrom\_chi2\_al}$ and $\allowbreak \code{astrom\_n\_good\_obs\_al}$ are the astrometric quantities available in the \Gaia DR2 catalog.

In addition, we removed nearby stars by making a parallax cut of $\varpi < 0.5 \mas$. 
We explored a more inclusive parallax cut that incorporated the uncertainty on the parallax measurement (similar to \citealt{Pace:2018}), but we found that such a cut increased contamination from faint foreground stars with large parallax  uncertainties.
We thus chose to retain our strict cut on parallax alone, though we recognize that such a cut will exclude some fainter members with larger parallax uncertainties.
This cut was applied for both analyses.

For flux measurements, we used the \SExtractor PSF magnitudes derived from the DES DR1 data. These magnitudes were corrected for interstellar reddening according to the procedure described in Section 4.2 of \citet{DES:2018}. 
We calculated a correction to the DES DR1 calibrated magnitudes in each band, $b$, according to $A_b = E(B − V) \times R_b$, where the fiducial coefficients were derived using the \citet{Fitzpatrick:1999} reddening law with $R_V = 3.1$ and the $E(B-V)$ values come from \citet[][]{Schlegel:1998}.
The coefficients $R_b$ were taken from \citet{DES:2018}: $R_g = 3.186$ and $R_r = 2.140$.
Throughout this paper, we use $g$ and $r$ to refer to the dereddened PSF magnitudes derived from DES DR1.
Our cross-matched sample ranges in magnitude from $16 \lesssim g \lesssim 21$, where the bright threshold is set by the saturation limit of DES and the faint threshold is set by the sensitivity of \Gaia.

For each stream, we transformed the data into a coordinate system aligned with the track of the stream, such that $\phi_1$ and $\phi_2$ are the coordinates along and across the track of the stream, respectively \citep[e.g.,][]{Koposov:2010}.
This transformation is performed by rotating the celestial equator to the great circle defined by the poles listed in Table 3 of \citet{Shipp:2018}, so that ($\phi_1, \phi_2$) = ($0^{\circ}, 0^{\circ}$) lies at the center of the observed portion of the stream. 
The rotation matrix for each stream is included in \appref{matrix}.

When analyzing each stream, we used proper motions converted into the rotated coordinate system and corrected for the solar reflex motion. We refer to these proper motions as $\mu_{\phi_1}, \mu_{\phi_2}$, where $\mu_{\phi_1}$ includes the $\cos \phi_2$ correction.
The velocity of the Sun relative to the Galactic standard of rest is taken to be $(U_{\odot}, V_{\odot}, W_{\odot}) = (11.1, 240.0, 7.3) \kms$ \citep{Schoenrich:2010,Bovy:2012} and we used the stream distances reported in \citet{Shipp:2018}. 

We then performed several data selections, some of which were applied generically to the data set, and some of which were applied selectively, depending on the stream and the analysis method. We enumerate these selection criteria below.

\begin{enumerate}[label=(\roman*)]

\item \label{itm:color_mag} \textbf{Color-magnitude: } 
We selected stars in $g-r$ vs.\ $g$ color-magnitude space following a method similar to that described in \citet{Pace:2018}. 
Red-giant branch (RGB) and main-sequence (MS) stars were selected based on the best-fit Dotter isochrones \citep{Dotter:2008} to the DES data. We began with the isochrone parameters listed in \citet{Shipp:2018}, then updated the age and metallicity of the isochrones based on the likely members after an iteration of the proper motion fit. The final isochrone values are listed in \tabref{isochrones}. 
For a given isochrone, we selected stars within either $\Delta (g-r) \pm 0.05$ mag or $\Delta g \pm 0.4$ mag of the Dotter isochrone.
In addition, we selected blue horizontal branch (BHB) stars using an empirical isochrone of M92 from \citet{Bernard:2014a} transformed to the DES photometric system. 
For the BHB selection, we used a wider color window, $\Delta (g-r) \pm 0.10$, 
due to the reduced foreground contamination at bluer colors. 
We did not select any red horizontal branch stars.

\item \label{itm:mag} \textbf{Magnitude: } In some cases, we made an additional magnitude cut that selected reasonably bright stars with smaller proper motion uncertainties. For the by-eye analysis of the brightest streams (i.e. ATLAS, Chenab, Jhelum, Phoenix, Ravi, Tucana III, Turranburra, and Wambelong), we selected stars with $g < 19$. In the Gaussian mixture model analysis, all streams had a cutoff at an absolute magnitude in the $g$-band of $M_g = 2$.

\item \label{itm:metal} \textbf{Metal-poor: } We performed a cut in $(g-r)$ vs.\ $(r-i)$ color-color space to select for metal-poor stars. Stars that lie along more metal-poor isochrones tend to have redder $r-i$ colors at a given $g-r$ color, as shown in \citet{Li:2018} and \citet{Pace:2018}. We selected stars that lie between 0.02 and 0.06 mag in $r - i$ above the empirical ridgeline of the stellar locus in DES. This selection was made only when necessary to further eliminate foreground contamination (i.e., Indus, Jhelum, Ravi, Turbio, Turranburra, and Wambelong).

\item \label{itm:spatial} \textbf{Spatial: } We selected a spatial region along each stream. For most streams, this is a region along the great circle connecting the stream's endpoints, as specified in \citet{Shipp:2018}. However, for ATLAS, which shows significant deviation from a great circle, the on-stream region was selected along the track defined by Equation~6 in \citet{Shipp:2018}. The widths of the on-stream selection varied between the two analysis methods. For the by-eye fit, we used the regions described in Table A.1 of \citet{Shipp:2018}. In contrast, for the Gaussian mixture model analysis, we define the on-stream region to be $3w$, where  $w$ represents the width of the stream, as reported in Table 1 of \citet{Shipp:2018}.

\item \label{itm:vesc} \textbf{Escape velocity: } When performing the GMM analysis (\secref{gmm}), we removed stars with tangential velocities greater than the Milky Way escape velocity at the distance of each stream.
We calculated the escape velocity, $v_{\rm esc}$, using the \code{MWPotential2014} from \code{galpy} \citep{Bovy:2015}, with a Milky Way halo mass of $M_{\rm vir} = 1.6 \times 10^{12}\Msun$, as in \citet{Pace:2018}. 
We calculated the tangential velocity, $v_{\rm tan}$, from the proper motion of each star, assuming the distance of the target stream, and removed all stars with $v_{\rm tan} > v_{\rm esc}$, in order to filter out nearby and possible hypervelocity stars. 
We verified that the analysis was robust against changes to this cut, e.g. by removing stars with $v_{\rm tan} - 3\sigma_{v_{\rm tan}} > v_{\rm esc}$.
\end{enumerate}

\section{Methods}
\label{sec:methods}

We obtained measurements of the proper motion of each stream with two complementary methods. First, we applied a set of physically-motivated cuts to increase the prominence of the stellar stream signal, which was estimated by eye based on the clustering of stars in proper-motion space. Second, we fit a GMM to the data to obtain a statistically robust measurement of the proper motion, proper motion gradient, and corresponding uncertainties for each stream. 
The by-eye fit yields a quick and intuitive measurement of the proper motion, while the GMM provides a more rigorous measurement including statistical uncertainties.

\subsection{By-Eye Fit}
\label{sec:eye}

\begin{figure*}
  \centering
    \includegraphics[width=0.75\textwidth]{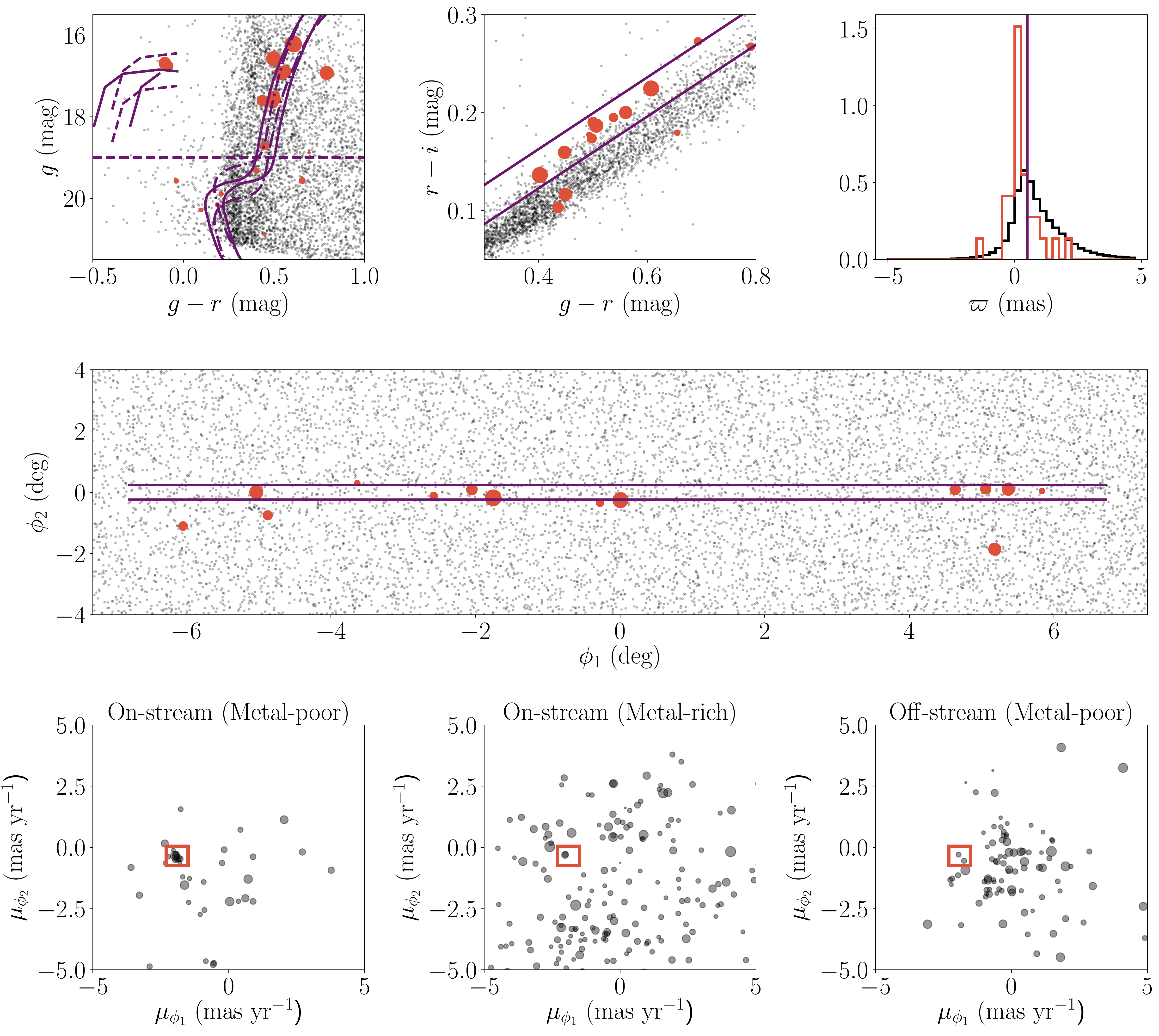}
    \caption{Example of the by-eye analysis for the Phoenix stream. The gray points illustrate the stars included in the cross-matched catalog surrounding the Phoenix stream. The purple lines indicate the selections made on color-magnitude (top left), color-color (top middle), parallax (top right), and spatial parameters (center). 
The lower three panels illustrate the proper motions of stars passing all selections (left), failing the metal poor cut (via color-color selection) but passing other selections (middle), and failing the on-stream cut but passing other selections (right).
A clear overdensity corresponding to the proper motion of the Phoenix stream (orange box) can be seen in the lower-left panel when all selections are applied.
Orange points in the color-magnitude and color-color panels reside within the orange box in proper motion and pass the parallax  and spatial selections.
The orange histogram in the parallax panel contains stars that reside within the orange box and pass the color-magnitude, color-color, and spatial sections.
Orange points in the center panel reside within the orange box and pass the color-magnitude, color-color, and parallax selections. The size of the orange points in the upper panels and the black points in the lower panels is inversely proportional to the uncertainty in the measured proper motion.}
    \label{fig:eye}
\end{figure*}

We applied a set of physically-motivated selection criteria to the data coincident with each stream to decrease foreground contamination and enhance the proper motion signature of stellar streams (enumerated in \secref{data}).
For all streams, we performed cuts on parallax, color-magnitude \ref{itm:color_mag}, and a color-color selection for metal-poor stars \ref{itm:metal}. 
We selected on-stream and off-stream regions for comparison with the local Milky Way foreground \ref{itm:spatial}. 
In addition, for a subset of bright streams (ATLAS, Chenab, Jhelum, Phoenix, Ravi, Tucana III, Turranburra, and Wambelong), we made a magnitude cut at $g < 19$ to remove stars with larger proper motion uncertainties \ref{itm:mag}. 
We visually inspected the cleaned data to identify clusters of stars in proper motion space that could correspond to the signatures of the stellar streams.
We identified possible proper motion signatures of thirteen streams (\tabref{results_eye}); nine of these are similarly identified by the GMM procedure described in \secref{gmm}.
Since the GMM procedure is more objective and statistically rigorous, we choose to report those values as our results; however, the by-eye fit proved to be a valuable diagnostic for validating the GMM fit.

\figref{eye} shows a graphical representation of these cuts applied to the Phoenix data. 
In the lower panels of \figref{eye}, the proper motions of stars passing three different selections are shown. The lower-left panel shows stars passing all selections, and the proper motion signal of the Phoenix stream is highlighted by the orange box. The lower-middle panel contains metal-rich stars that lie along the stream, and the lower-right panel shows metal-poor stars in the off-stream region. As expected, the proper motion signal is only visible in metal-poor stars that lie along the track of the Phoenix stream. Stars with a proper motion consistent with our measurement of the Phoenix stream (within the orange box) are plotted in the other panels of Figure \ref{fig:eye} and are found to be consistent with the Phoenix stream in color, magnitude, and location on the sky.

\subsection{Gaussian Mixture Model Fit}
\label{sec:gmm}

\begin{figure*}
  \centering
    \includegraphics[width=0.95\textwidth]{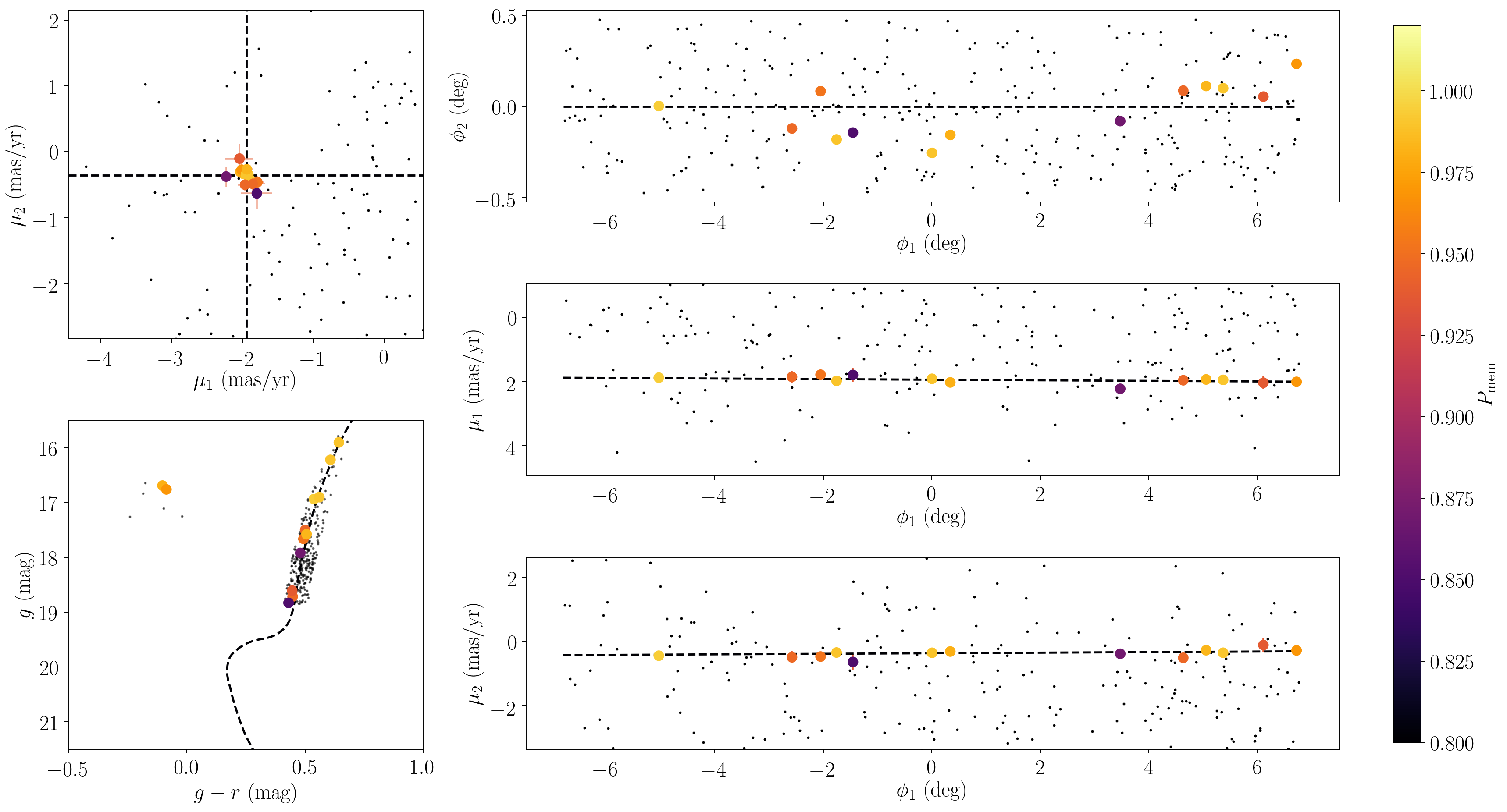}
    \caption{Results of the Gaussian mixture model fit to the Phoenix stream. The black points illustrate the data that was included in the fit; stars with membership probabilities $P_{\rm mem} > 0.8$ are color-coded by their membership probabilities.}
    \label{fig:gmm}
\end{figure*}

After obtaining measurements by eye, we fit a Gaussian mixture model to the data. We performed the fit on all 14 streams, and obtained results consistent with the by-eye method for the nine confidently-measured streams.
The fitting procedure follows that of \citet{Pace:2018} and is described briefly below.

The mixture model includes two multivariate Gaussian components in proper motion space. The first describes the stream, and has a dispersion fixed to zero.
The second component accounts for the Milky Way foreground and any other contaminating stellar populations. 
For each stream, we also tried introducing a third component to test whether the background would be better described by two Gaussians and in no case were the resulting stream parameters affected by the additional component. We therefore model the background by a single Gaussian component in the results presented here.

The likelihood is calculated as a product of two parts, the proper motion distribution and the spatial distribution. The proper motion term is modeled as,
\begin{equation}
\ln \mathcal{L}_{\rm PM} = -\frac{1}{2} (\chi - \overline{\chi})^\top C^{-1} (\chi - \overline{\chi}) - \frac{1}{2}\ln{\left(4 \pi^2 \det C \right)},
\end{equation}
\noindent where $\chi = ( \mu_{\phi_1}, \mu_{\phi_2})$ is the data vector and $\overline{\chi} = ( \overline{\mu_{\phi_1}} (\phi_1), \overline{\mu_{\phi_2}} (\phi_1))$ is the vector containing the systemic proper motion of the stream as a function of $\phi_1$. Allowing the systemic proper motion to vary with $\phi_1$ gives us a measurement of the proper motion gradient along the length of the stream. The covariance matrix, $C$, includes the correlation between the proper motion errors and a term for the intrinsic proper motion dispersion.  The covariance matrix is:

\begin{equation}
C=
\begin{bmatrix}
\epsilon_{\mu_{\rm \phi_1}}^2 +  \sigma_{\rm \mu_{\phi_1}}^2 & \epsilon_{\rm [\mu_{\phi_1} \times \mu_{\rm \phi_2}]}^2 \\
\epsilon_{[\mu_{\rm \phi_1} \times \mu_{\rm \phi_2}]}^2 & \epsilon_{\rm \mu_{\phi_2}}^2 +  \sigma_{\rm \mu_{\rm \phi_2}}^2  
\end{bmatrix} 
\, . \\ 
\end{equation}

The second part of the mixture model is a spatial prior based on the distance between stars and the stream axis in $\phi_2$.  We included the spatial stream prior probability as a truncated normal distribution, with a standard deviation equal to the stream width listed in Table 1 of \citet{Shipp:2018}.  For the Milky Way, the spatial prior probability was approximated as a uniform distribution across the narrow region included in the fit.
The complete set of free parameters and their priors are listed in Table \ref{table:priors}.

\newcommand{\priorcaption}{Priors on Gaussian Mixture Model.}
\newcommand{\priorcomments}{Priors on the nine free parameters in the Gaussian mixture model fits. $\mu_{\rm \phi_{1,2}}$ are the mean stream proper motions at $\phi_1$ = 0. $d \mu_{\rm \phi_{1,2}}/d \phi_1$ are the gradients of the stream proper motions with respect to $\phi_1$. $\mu_{\rm \phi_{1,2},\mathrm{MW}}$ are the mean proper motions of the Milky Way foreground component. $\sigma_{\rm \mu_{\rm \phi_{1,2}},\mathrm{MW}}$ are the dispersions of the Milky Way proper motions. $f_{\rm MW}$ is the fraction of stars belonging to the Milky Way component.}
\begin{deluxetable}{l ccc c}
\tablecolumns{13}
\tablewidth{0pt}
\tabletypesize{\scriptsize}
\tablecaption{ \priorcaption }
\tablehead{Parameter~~~~~~ & ~~~~~Prior~~~~~ & ~~~~~Range~~~~~ & ~~~Units~~~}
\startdata
$\mu_{\rm \phi_{1,2}}$                      & Uniform  & (-10, 10) & mas/yr \\
$d \mu_{\rm \phi_{1,2}}/d \phi_1$           & Uniform  & (-3, 3)   & mas/yr/deg \\
$\mu_{\rm \phi_{1,2},\mathrm{MW}}$          & Uniform  & (-10, 10) & mas/yr \\
$\sigma_{\rm \mu_{\rm \phi_{1,2}},\mathrm{MW}}$ & Jeffreys & (-1, 3) & mas/yr \\
$f_{\rm MW}$                        & Uniform  & (0, 1) & \\
\enddata
{\footnotesize \tablecomments{ \priorcomments }}
\label{table:priors}
\end{deluxetable}


Before performing the mixture model fit, we first made several data selections as described in \secref{data}. For all streams, we made cuts on parallax and astrometric fit quality. Cuts on color-magnitude \ref{itm:color_mag} and escape velocity \ref{itm:vesc} were made for each stream individually. Several thicker streams required additional filtering, so we performed the metal-poor selection \ref{itm:metal} on Indus, Jhelum, Ravi, Turbio, Turranburra, and Wambelong.

Following \citet{Pace:2018}, we use the \code{MultiNest} algorithm \citep{Feroz:2008, Feroz:2009} to determine the posterior distribution. We compute a Bayes factor to assess the significance of each stream signal, comparing models with only the Milky Way component, and with both the Milky Way and stream components.

As an example, we show the results of the Gaussian mixture model fit to Phoenix in Figure \ref{fig:gmm}. All stars included in the fit are plotted, with stars with $P_{\rm mem} > 0.8$ colored by their membership probability.

\newcommand{\derivcaption}{Derived proper motion of DES streams.}
\newcommand{\derivcomments}{The first two columns are proper motion measurements in the observed coordinate system. Fits to Indus and Jhelum did not converge without first correcting for the solar reflex motion. All uncertainties reported here are statistical uncertainties from the mixture model fitting. Additional uncertainties, including the uncertainty propagated from the distance measurement through the correction for the solar reflex motion, are not included. We find that Jhelum is best fit by a two-stream-component model. The first row lists the result of fitting a single stream component to Jhelum, and Jhelum-a and Jhelum-b are the results of each component from the two-component fit. The tangential velocity is calculated by $v_{\rm tan} = 4.74 d \mu$ km/s, where $d$ is the distance measured in kpc, and $\mu$ is the proper motion measured in mas/yr.}
\begin{deluxetable*}{l ccccccccc }[ht]
\tablecolumns{13}
\tablewidth{0pt}
\tabletypesize{\scriptsize}
\tablecaption{ \derivcaption }
\tablehead{ & $\mu_{\rm \alpha} cos \delta$ & $\mu_{\rm \delta}$ & $\mu_{\rm \phi_1}$ & $\mu_{\rm \phi_2}$ & $d \mu_{\rm \phi_1}/d \phi_1$ & $d \mu_{\rm \phi_2}/d \phi_1$ & $v_{\rm tan}$ & Bayes Factor \\
 & (mas/yr) & (mas/yr) & (mas/yr) & (mas/yr) & (mas/yr/deg) & (mas/yr/deg) & (km/s) & &  }
\startdata
Aliqa Uma       & $ 0.25 \pm 0.03$ & $-0.71 \pm 0.05$ & $ 0.98 \pm 0.04$ & $-0.34 \pm 0.03$ & $-0.02 \pm 0.03$ &  $-0.04 \pm 0.02 $ & 141 & -2.3   \\
ATLAS           & $ 0.09 \pm 0.03$ & $-0.88 \pm 0.03$ & $ 1.66 \pm 0.04$ & $-0.15 \pm 0.05$ & $ 0.02 \pm 0.005$ &  $-0.02 \pm 0.005 $ & 181 & 18.0   \\
Chenab          & $ 0.32 \pm 0.03$ & $-2.47 \pm 0.04$ & $ 1.03 \pm 0.05$ & $-0.60 \pm 0.03$ & $ 0.04 \pm 0.01$ &  $-0.02 \pm 0.01 $ & 225 & 15.2   \\
Elqui           & $ 0.13 \pm 0.04$ & $-0.33 \pm 0.09$ & $ 0.56 \pm 0.06$ & $-0.03 \pm 0.05$ & $-0.03 \pm 0.02$ &  $-0.04 \pm 0.01 $ & 133 & 13.2   \\
Indus           &         --       &         --       & $-3.09 \pm 0.03$ & $ 0.21 \pm 0.03$ & $ 0.05 \pm 0.004 $ &  $ 0.04 \pm 0.004 $ & 245 & 15.5   \\
Jhelum          &         --       &         --       & $-6.00 \pm 0.03$ & $-0.83 \pm 0.05$ &         --       &          --        & 378 & 9.4   \\
~~Jhelum-a      &         --       &         --       & $-6.01 \pm 0.02$ & $-0.84 \pm 0.04$ &         --       &          --        & 379 & \multirow{2}{*}{22.9}  \\
~~Jhelum-b      &         --       &         --       & $-4.97 \pm 0.03$ & $ 0.11 \pm 0.06$ &         --       &          --        & 310 &  \\
Phoenix         & $ 2.76 \pm 0.02$ & $-0.05 \pm 0.02$ & $-1.94 \pm 0.02$ & $-0.36 \pm 0.02$ & $-0.01 \pm 0.01$ & $  0.01 \pm 0.01 $ & 179 & 7.8    \\
Tucana III      & $-0.10 \pm 0.04$ & $-1.64 \pm 0.04$ & $ 1.08 \pm 0.03$ & $-0.03 \pm 0.03$ & $ 0.12 \pm 0.03$ & $ -0.06 \pm 0.03 $ & 129 & 28.2   \\
Turranburra     & $ 0.43 \pm 0.04$ & $-0.89 \pm 0.04$ & $ 0.69 \pm 0.04$ & $-0.22 \pm 0.04$ & $ 0.00 \pm 0.02$ & $ -0.03 \pm 0.01 $ &  95 & 2.7    \\
\enddata
{\footnotesize \tablecomments{ \derivcomments }}
\label{tab:results}
\end{deluxetable*}

\section{Results}
\label{sec:results}

\begin{figure*}
    \centering
    \gridline{\fig{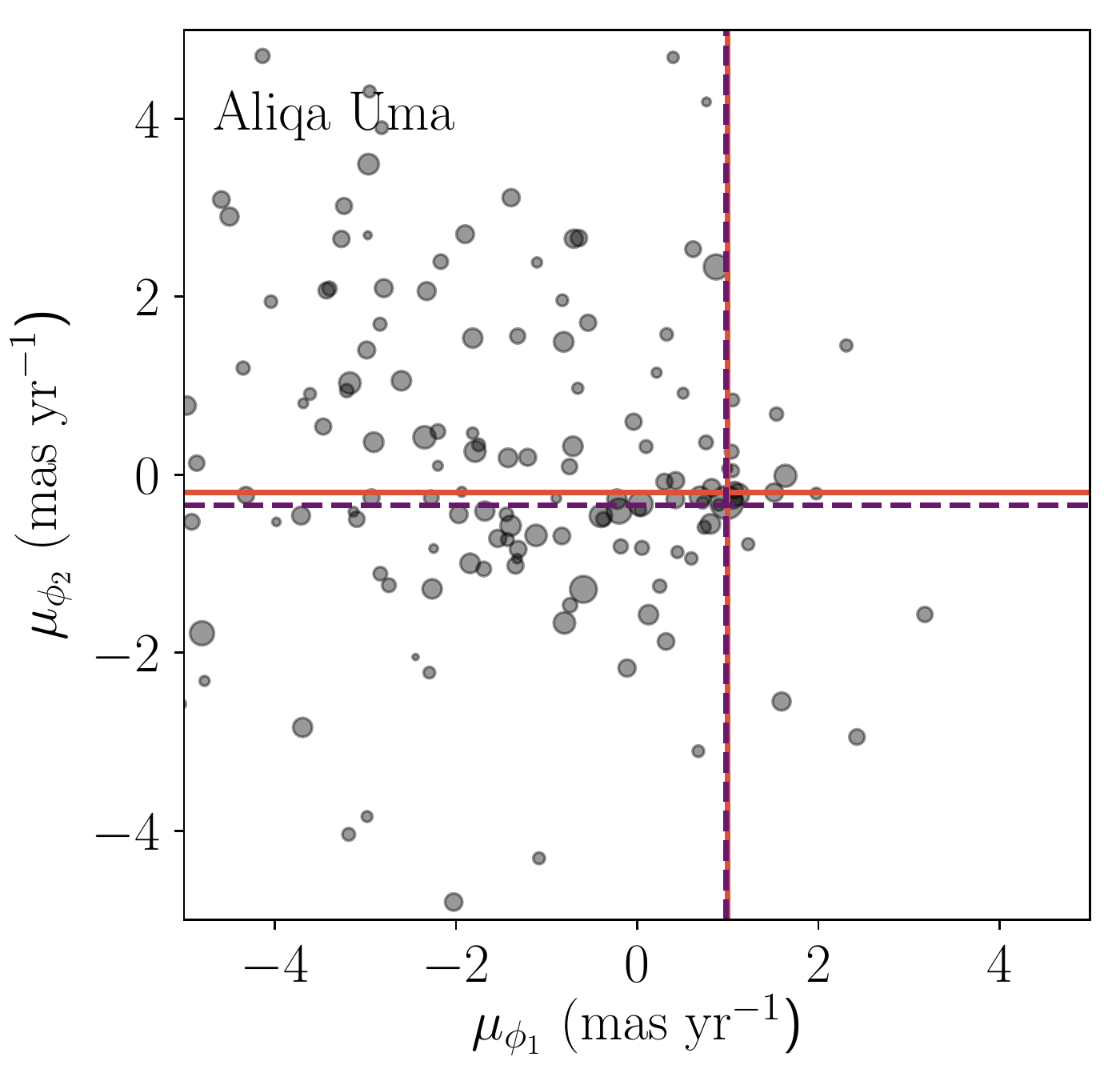}{0.3\textwidth}{}
              \fig{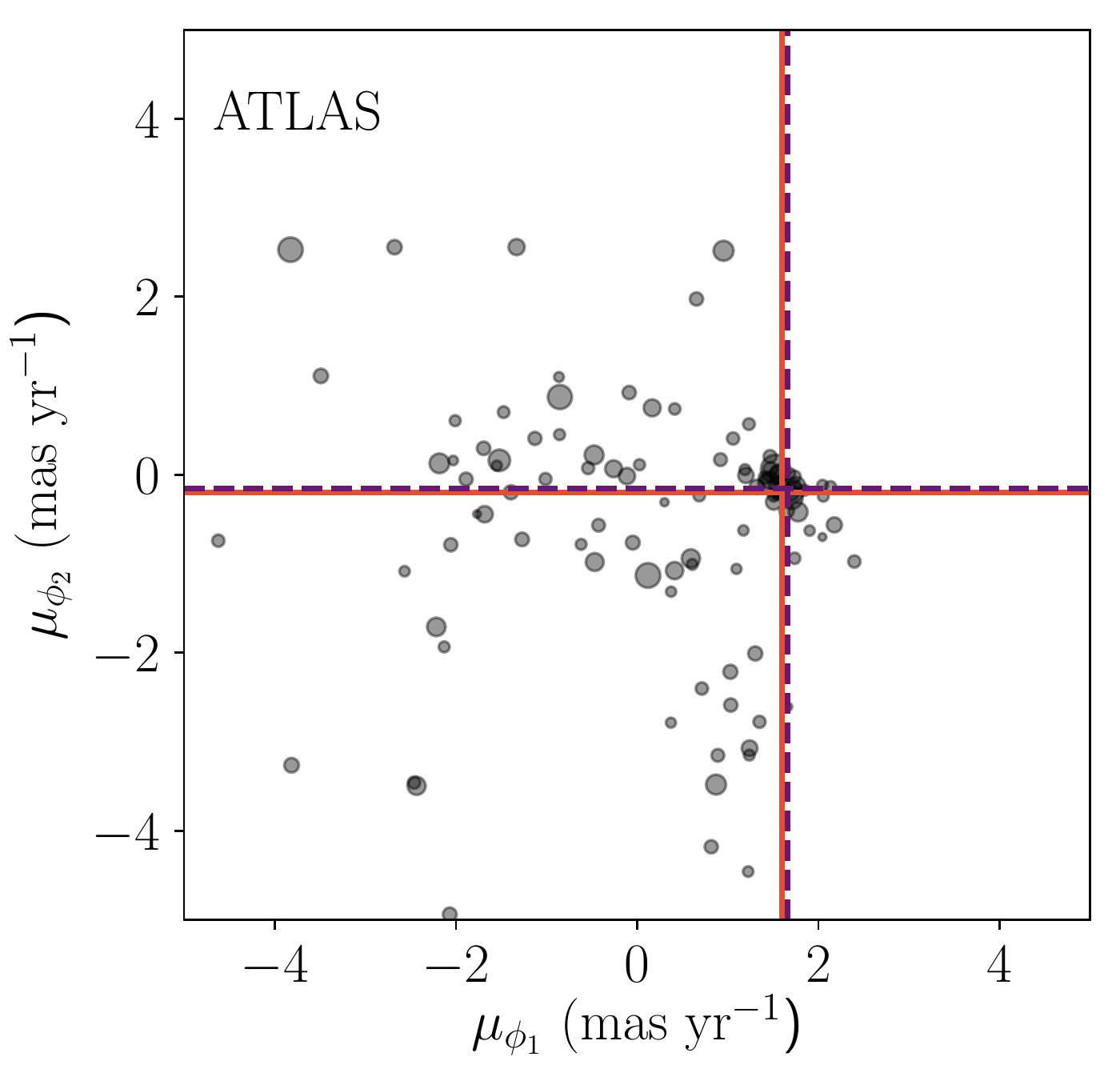}{0.3\textwidth}{}
              \fig{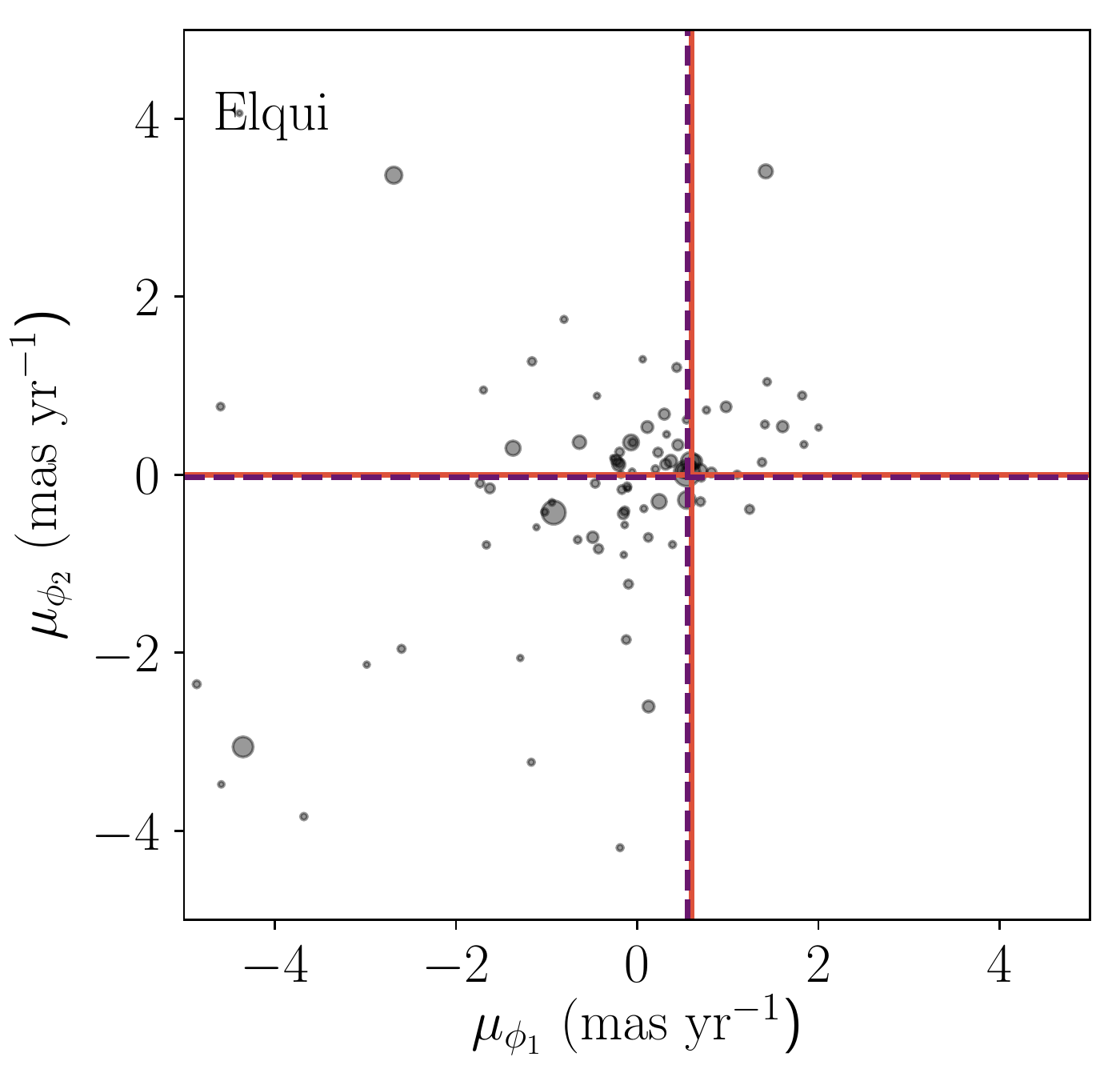}{0.3\textwidth}{}}
    \gridline{\fig{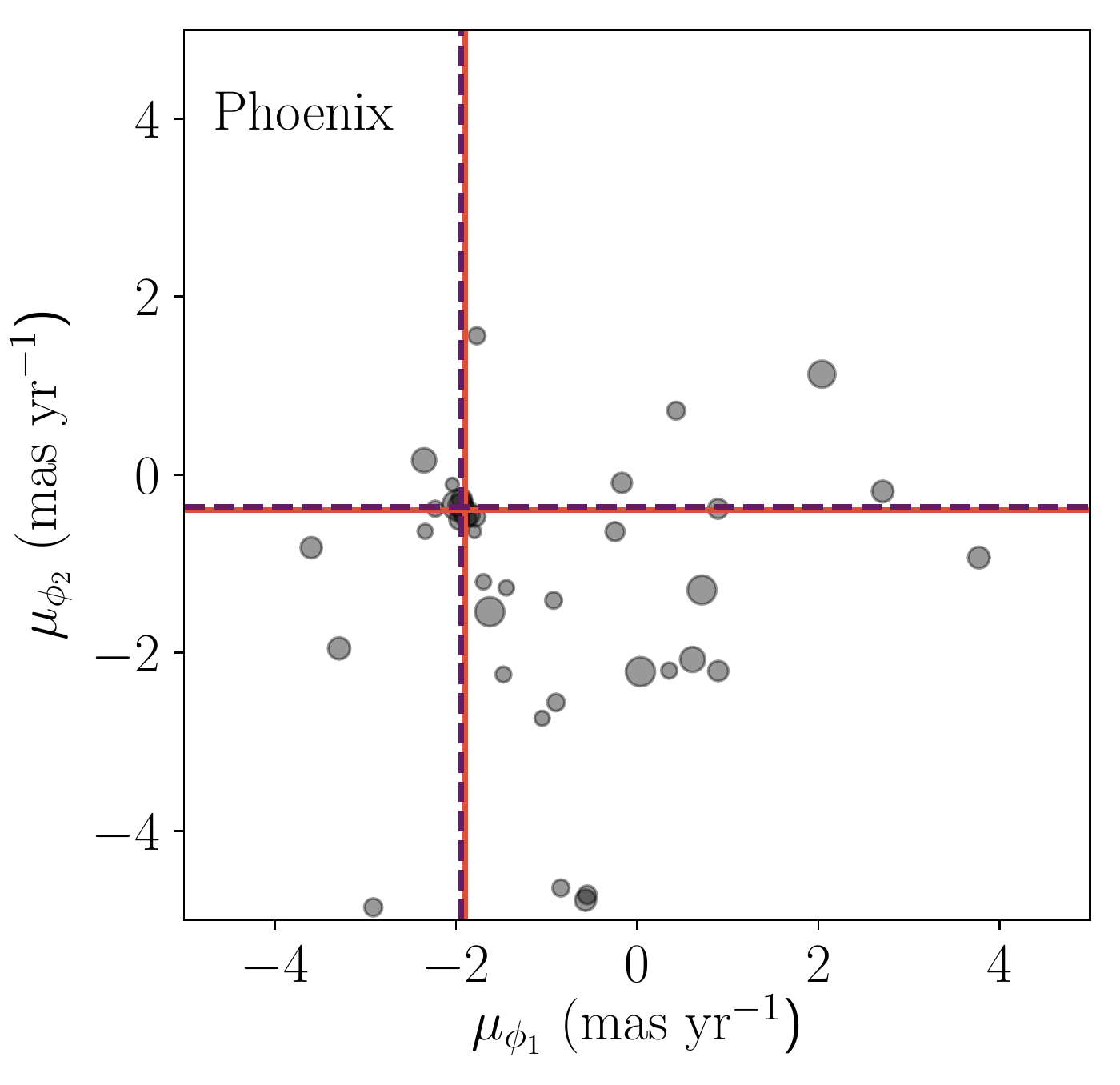}{0.3\textwidth}{}
              \fig{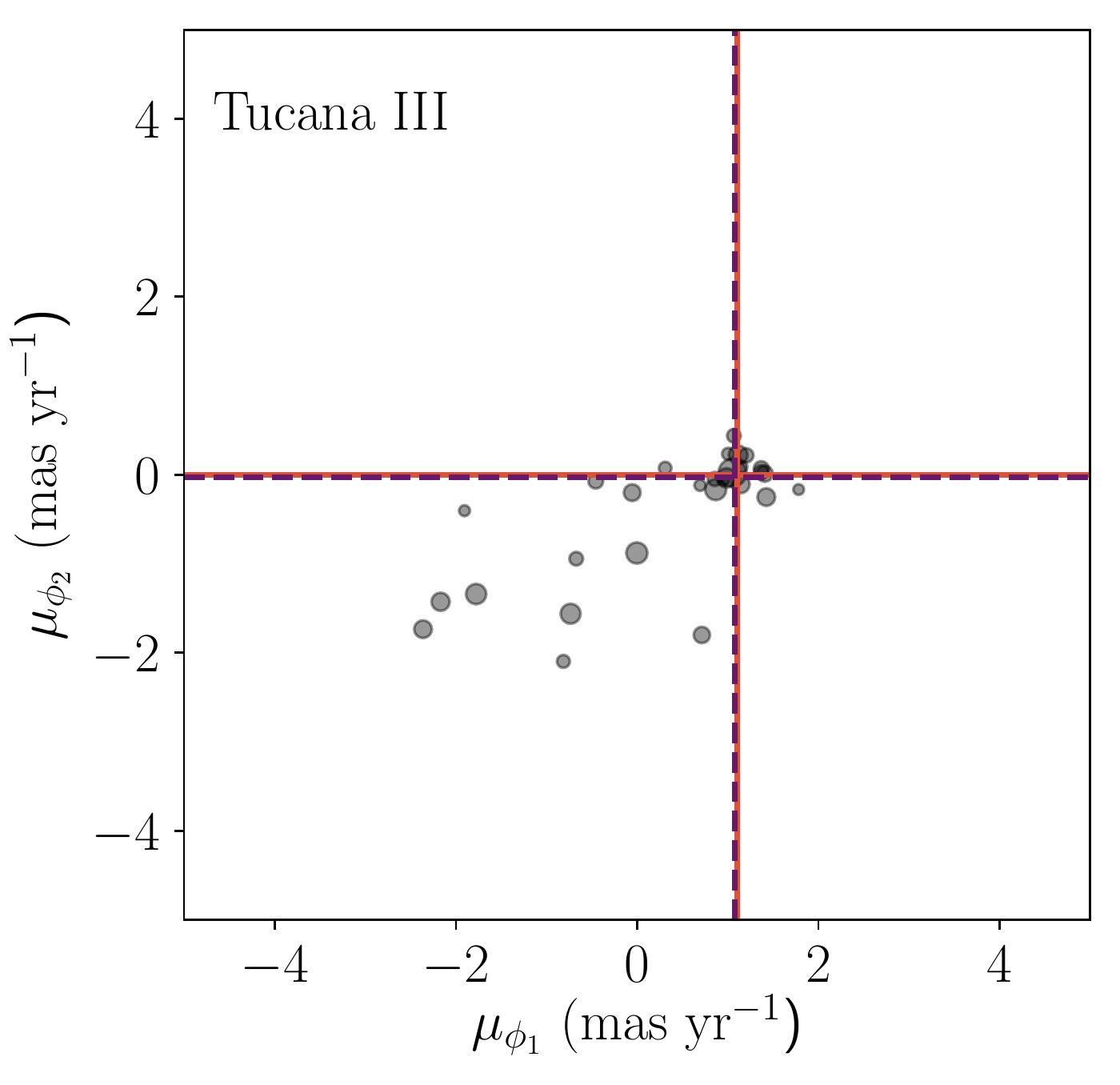}{0.3\textwidth}{}
              \fig{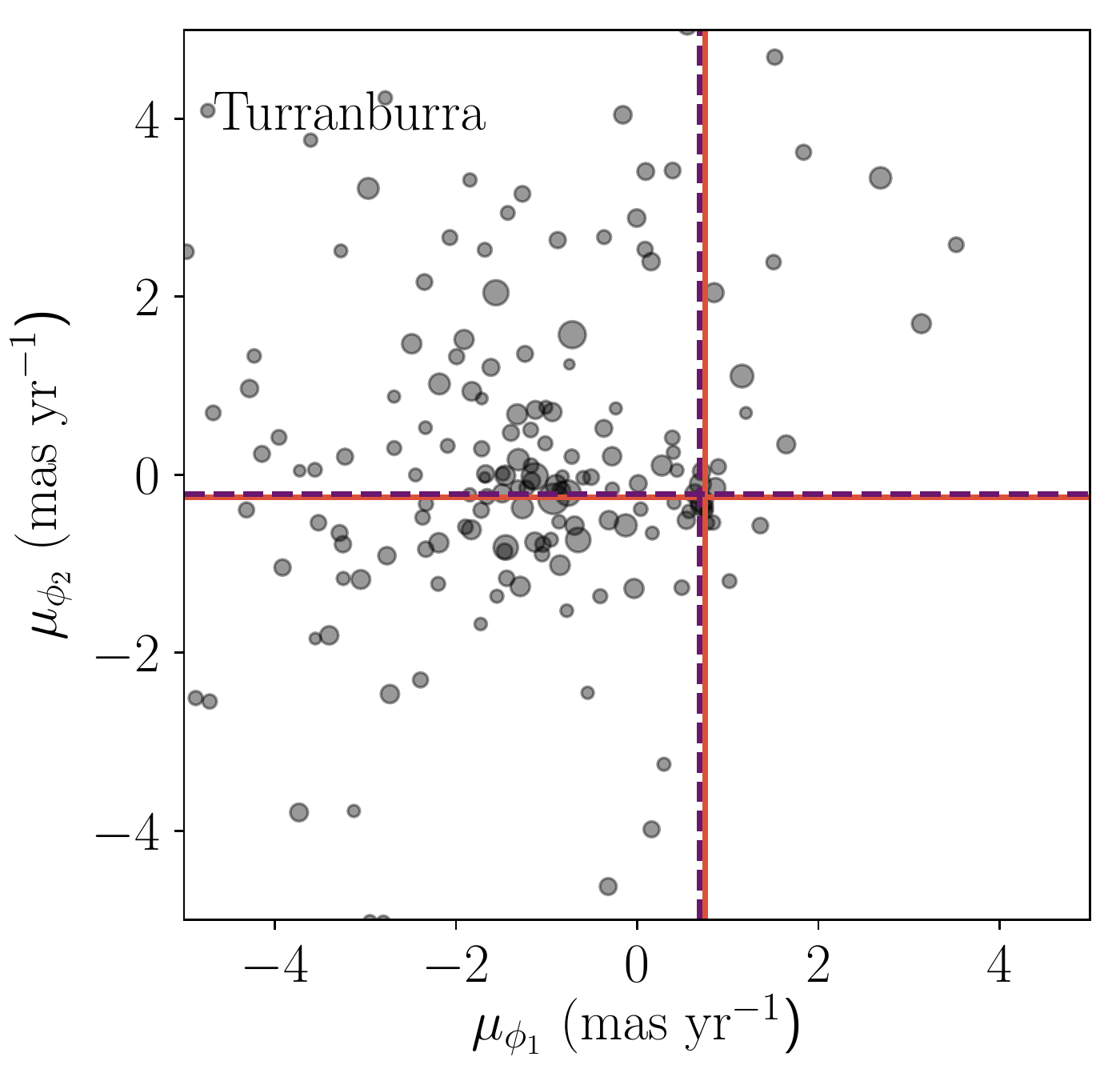}{0.3\textwidth}{}}
    \gridline{\fig{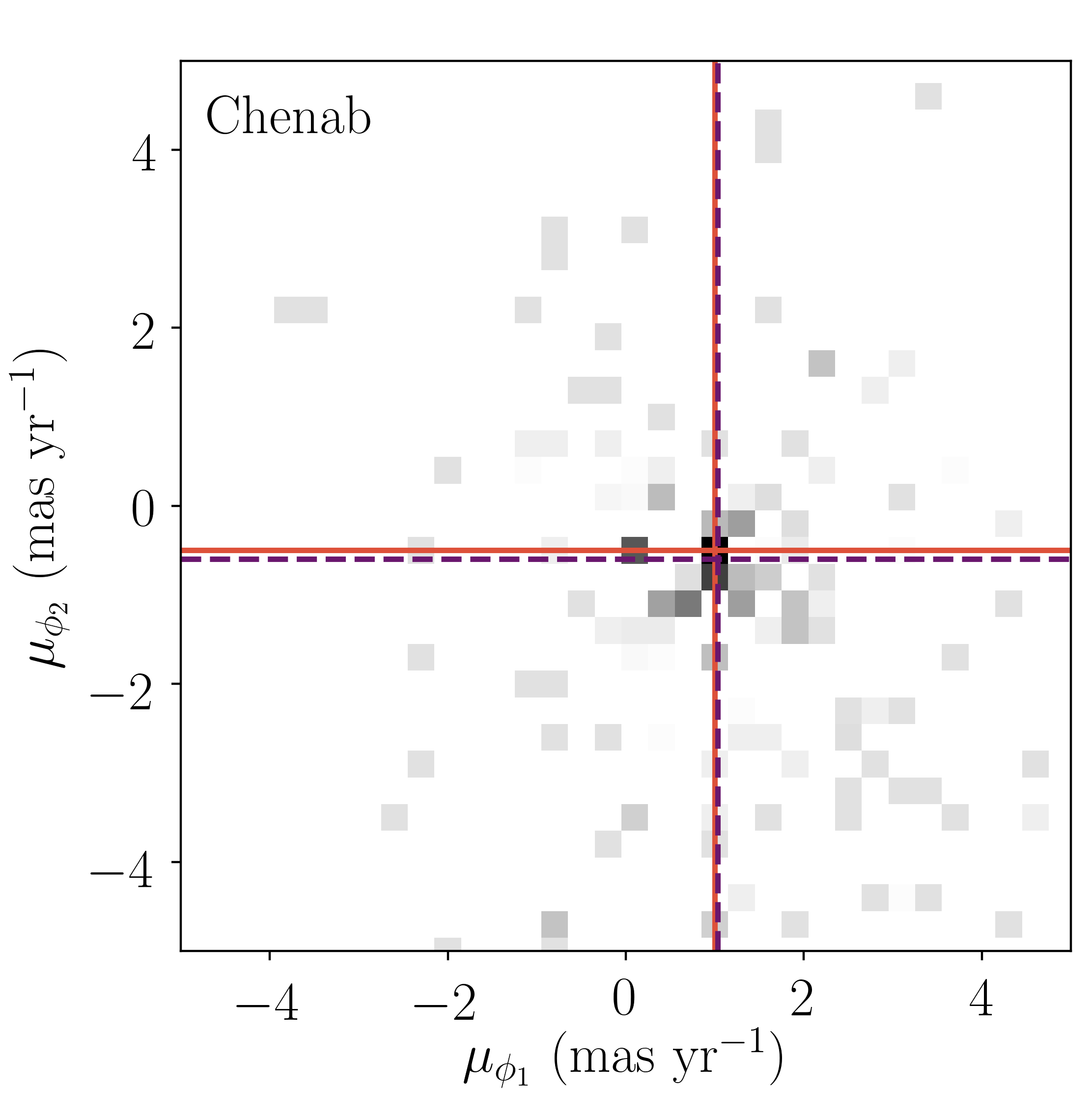}{0.3\textwidth}{}
              \fig{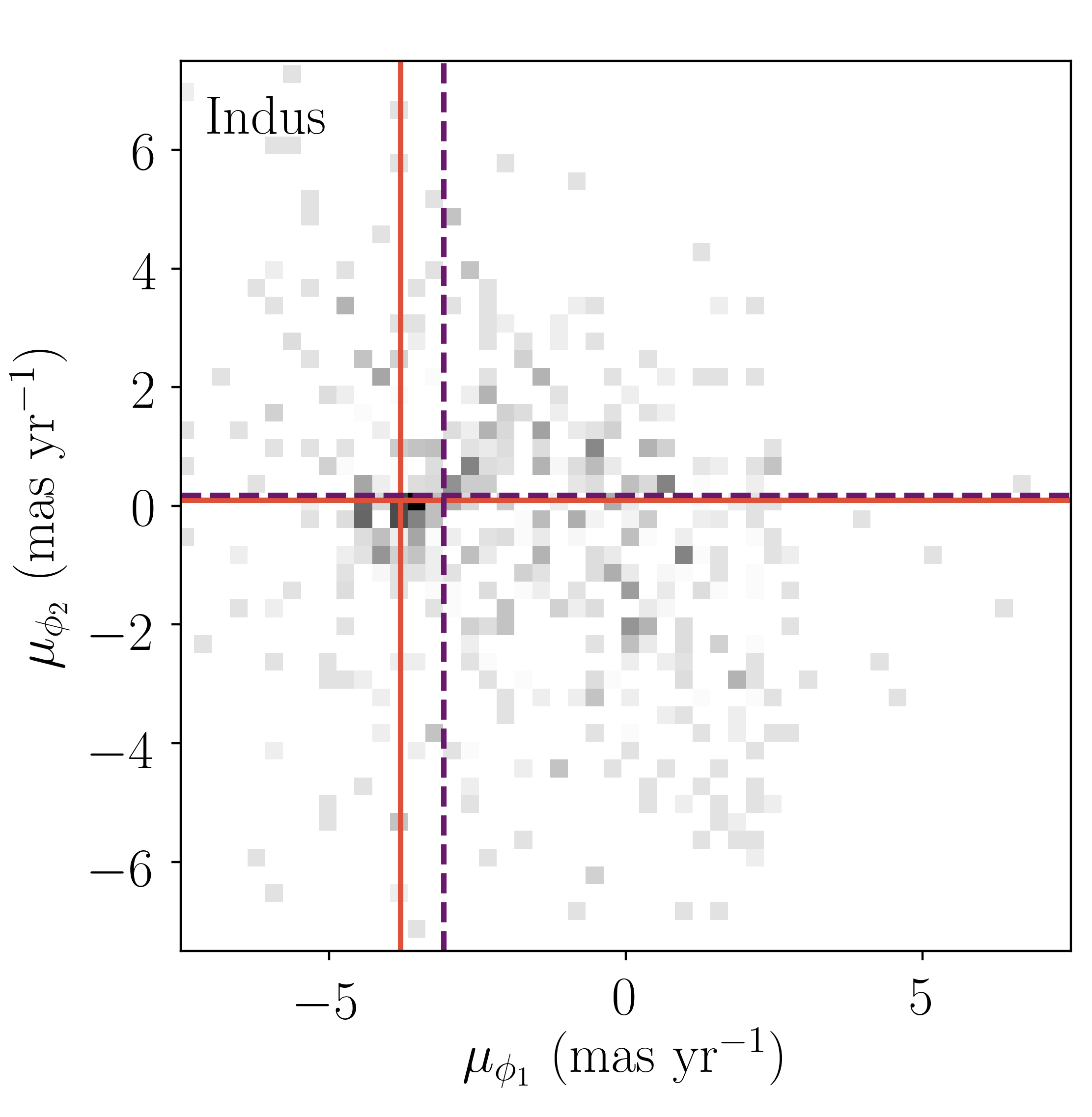}{0.3\textwidth}{}
              \fig{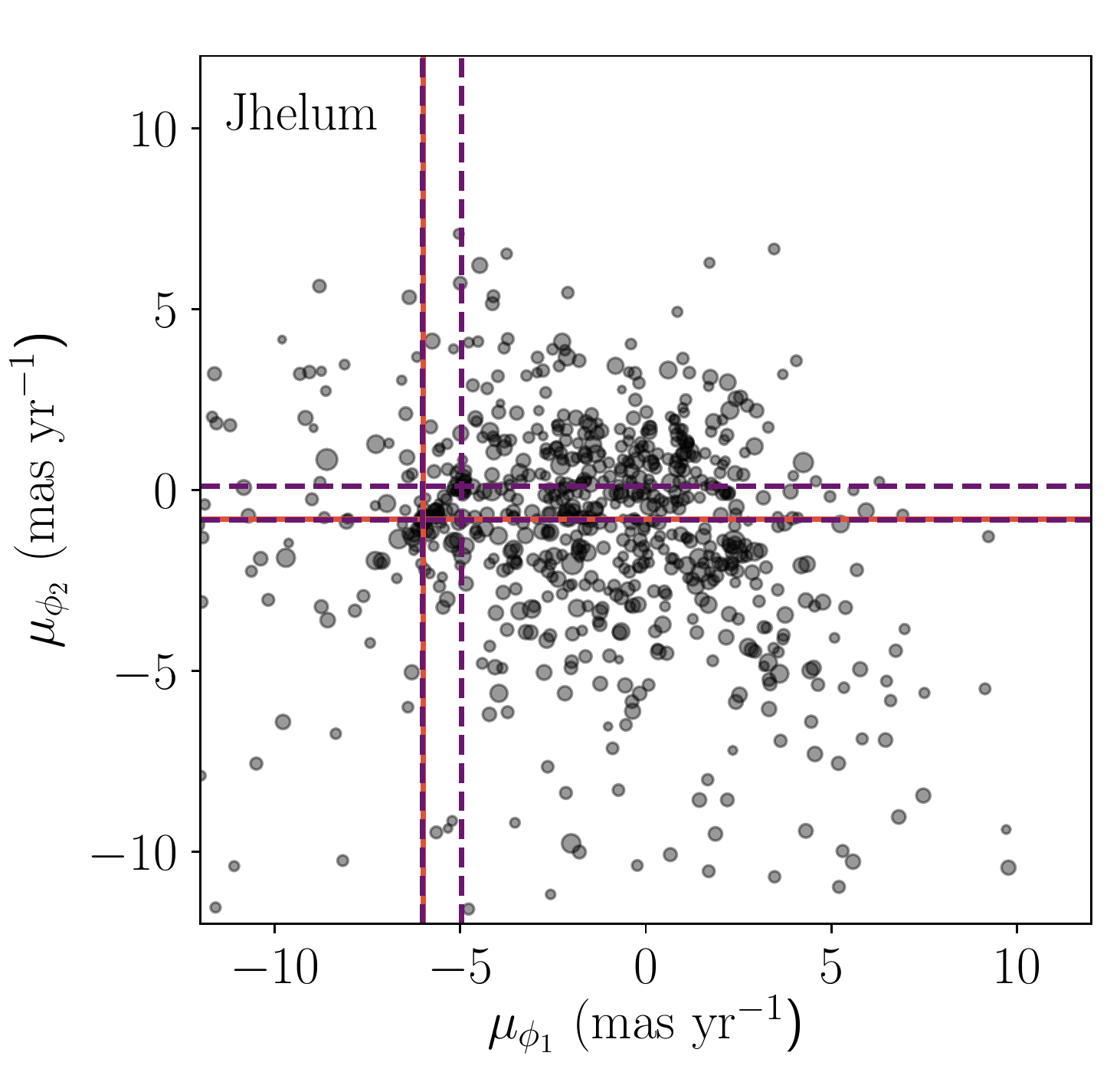}{0.3\textwidth}{}}
    \caption{{\Gaia} DR2 proper motion measurements of confidently detected streams in the DES footprint. 
    Proper motions are transformed to stream coordinates, $\mu_{\phi_1}, \mu_{\phi_2}$.
    Best-fit proper motion estimates fit by eye are shown by the solid orange crosshairs, while the best-fit results from the GMM are shown in purple.
    In the scatter plots, the size of the points is inversely proportional to the $1\sigma$ uncertainty on the proper motion of each star. Chenab and Indus, two of the thicker streams, are better shown by density histograms.}
    \label{fig:eye_signal}
\end{figure*}

\begin{figure*}
    \centering
    \gridline{\fig{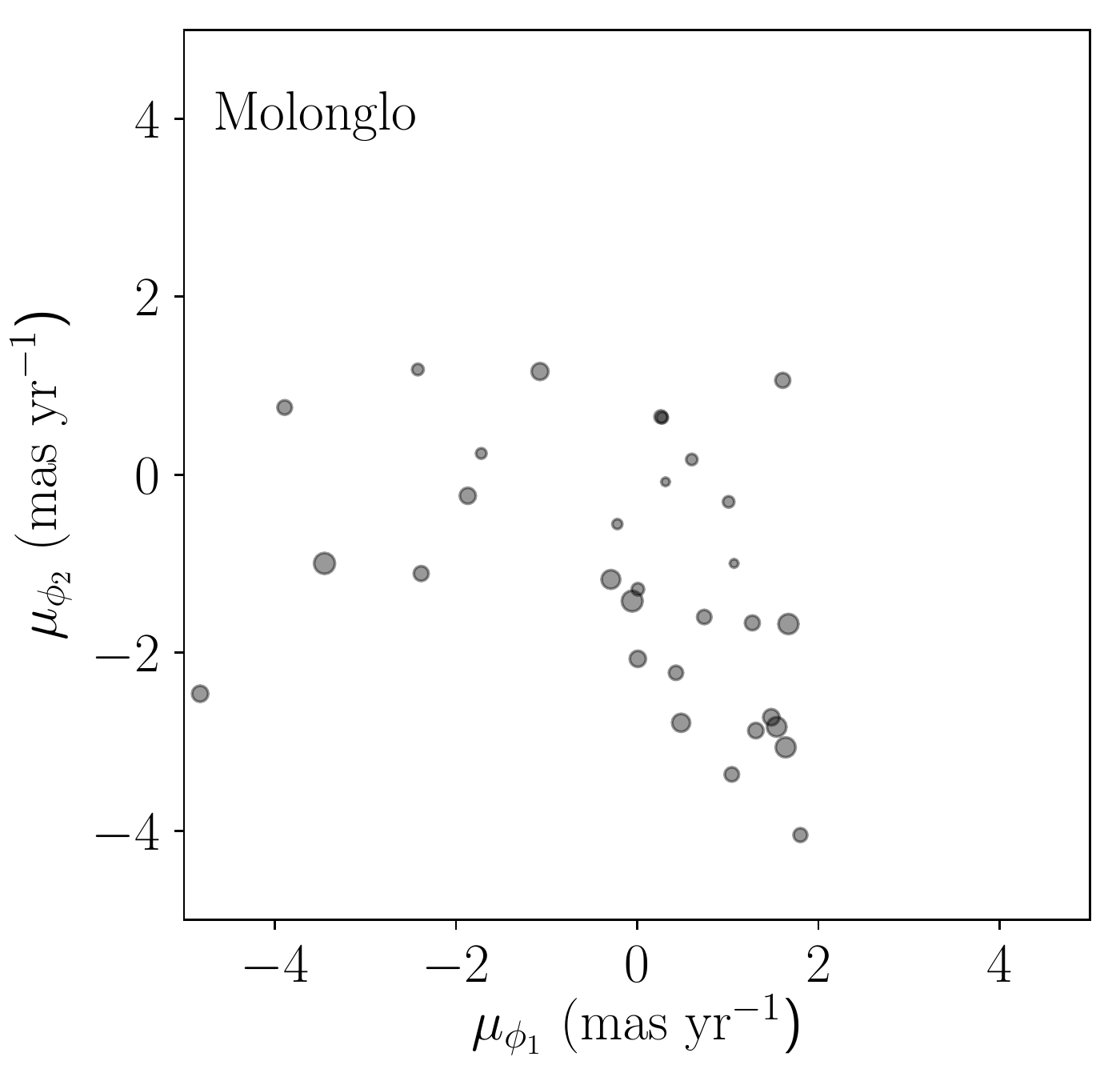}{0.3\textwidth}{}
              \fig{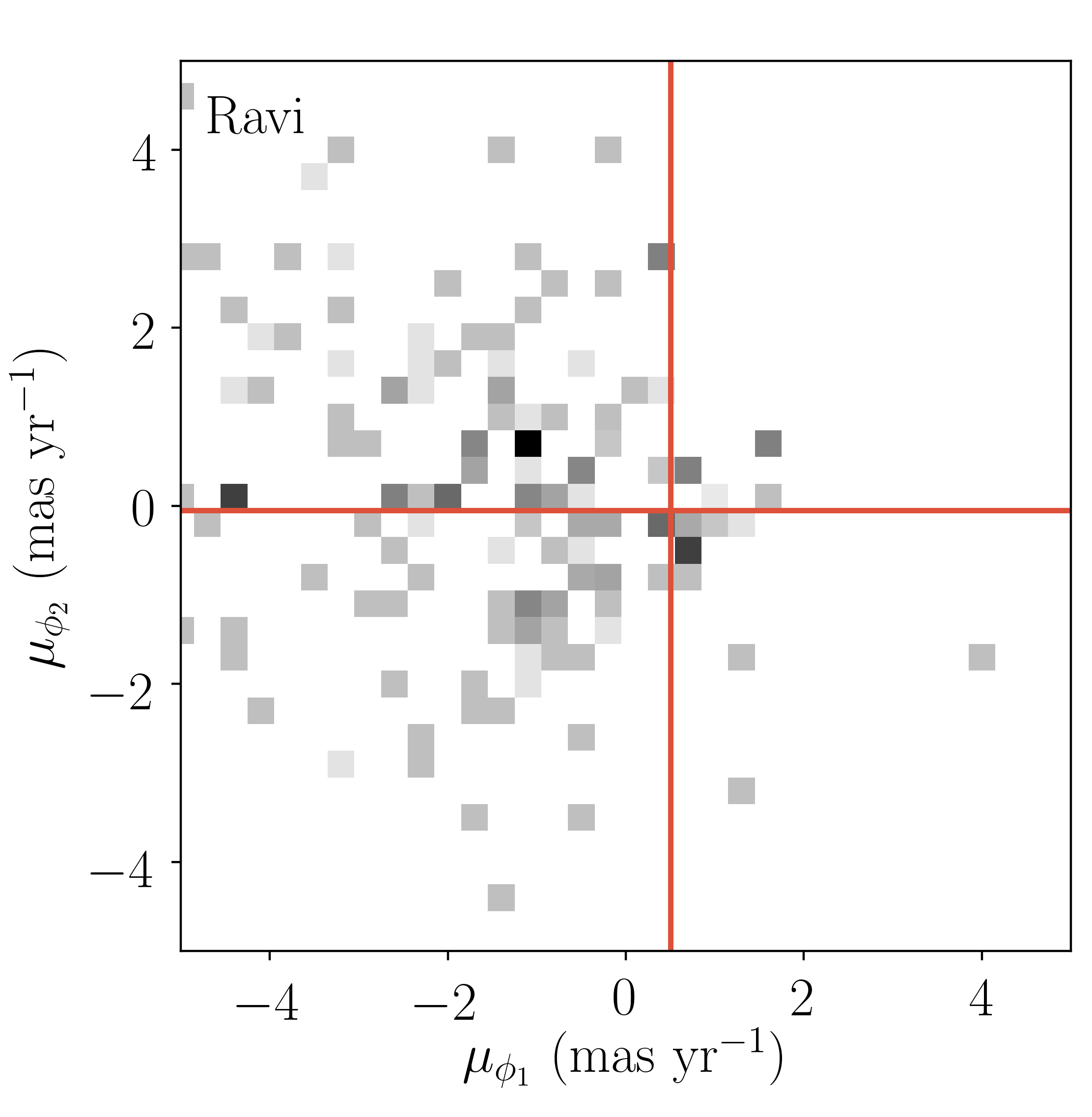}{0.3\textwidth}{}
              \fig{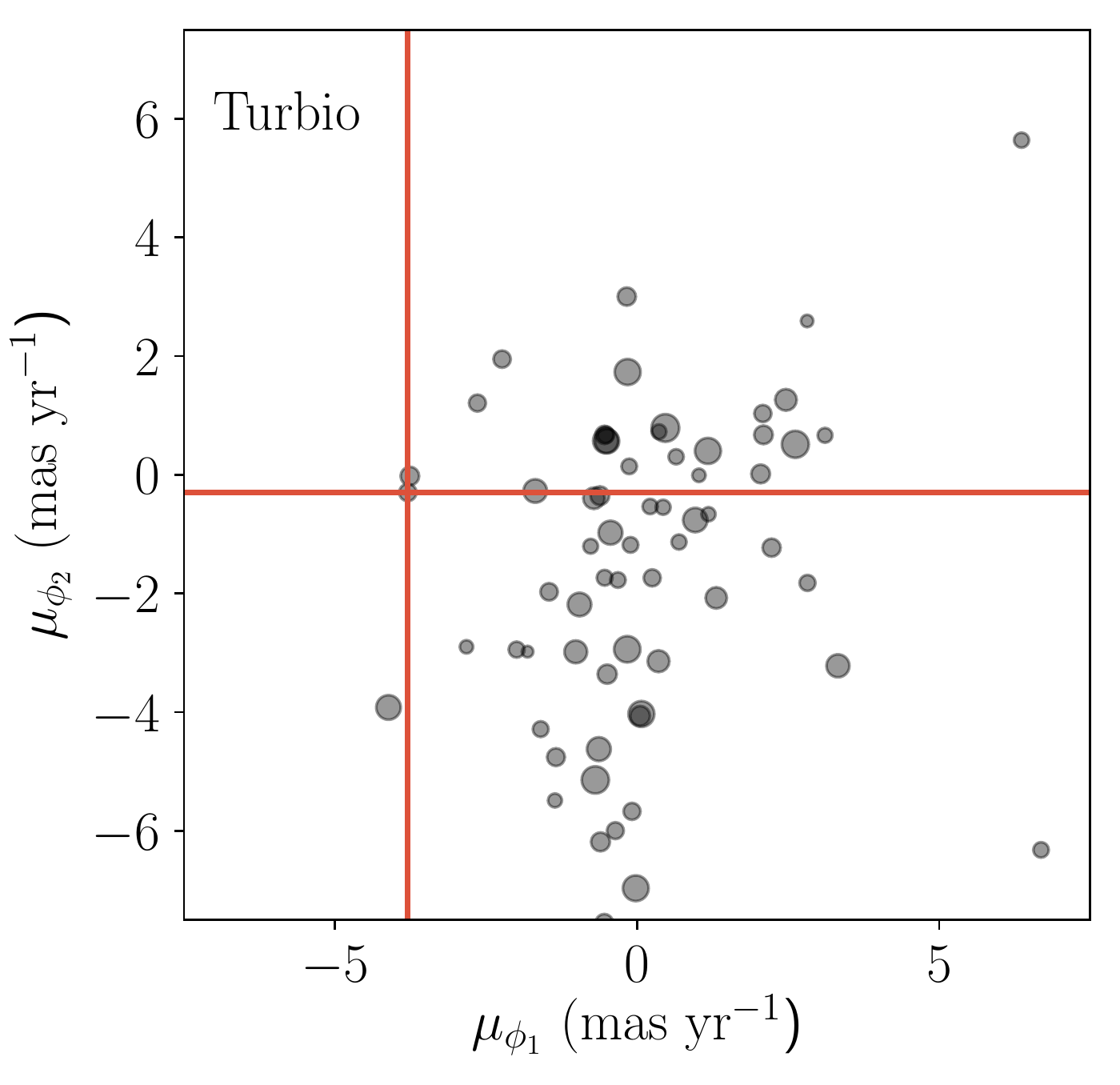}{0.3\textwidth}{}}
    \gridline{\fig{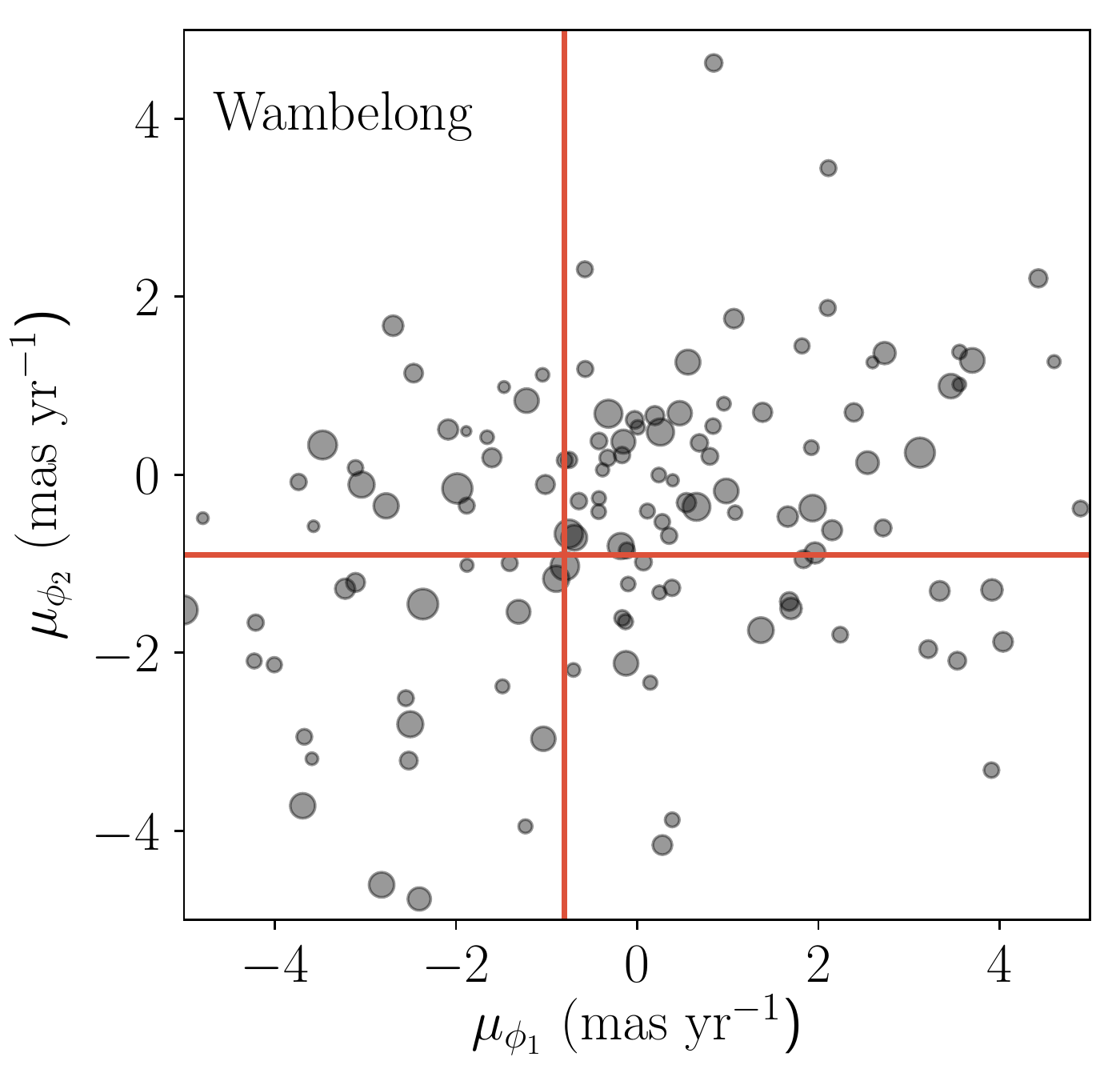}{0.3\textwidth}{}
              \fig{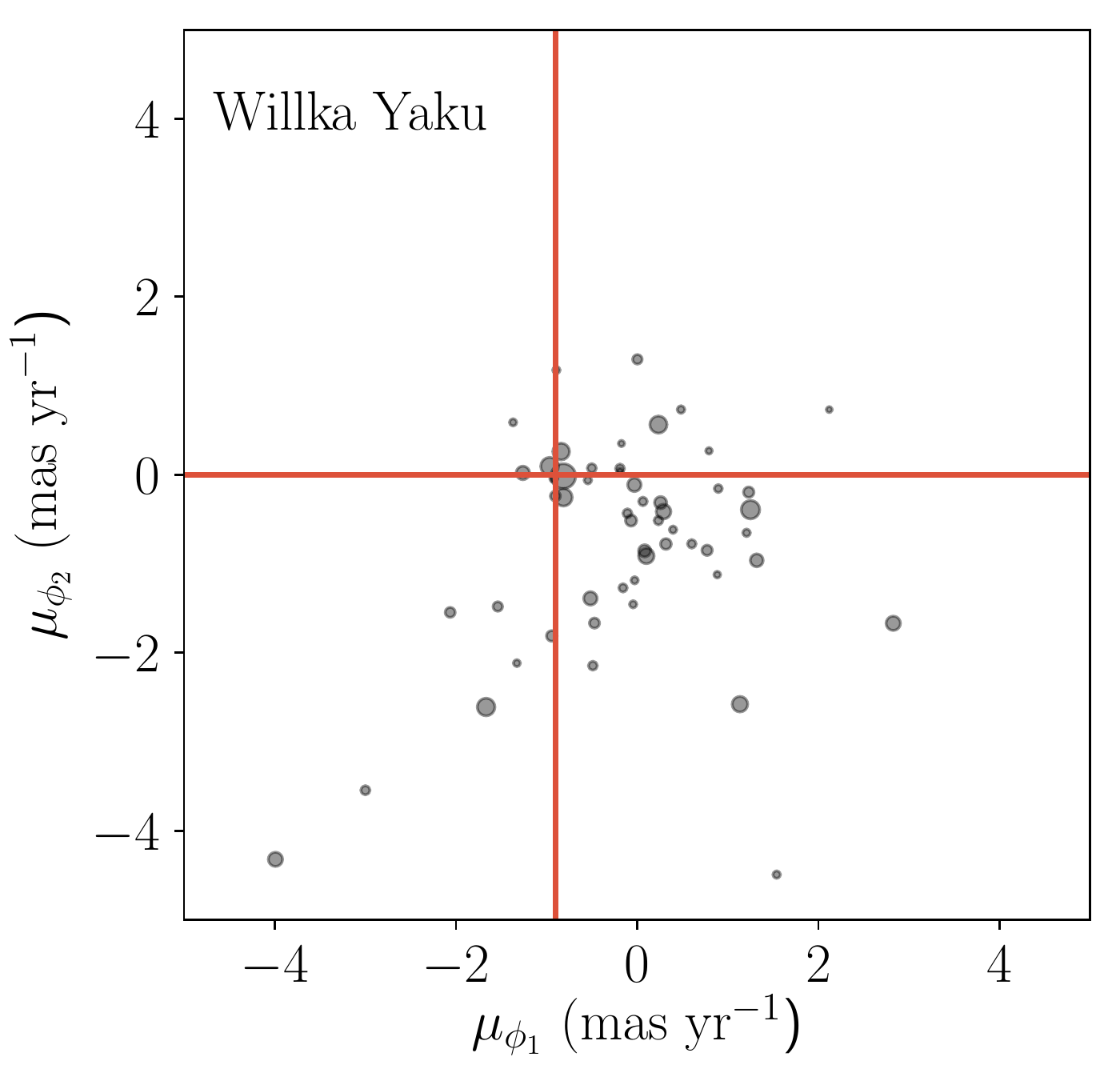}{0.3\textwidth}{}}
    \caption{{\Gaia} DR2 proper motion measurements for low-detection-confidence streams in the DES footprint. Low-confidence by-eye proper motion measurements are shown by the orange crosshairs.}
    \label{fig:eye_nosignal}
\end{figure*}

The by-eye and GMM analyses yield proper motion measurements for nine streams: Aliqa Uma, ATLAS, Chenab, Elqui, Indus, Phoenix, Jhelum, Tucana III, and Turranburra, as illustrated in \figref{eye_signal}. 
In the Figure, the points indicate stars included in the by-eye analysis. 
The solid crosshairs indicate the by-eye measurement, and the dashed crosshairs mark the result of the GMM fit. 
The by-eye and GMM measurements agree quite well for the majority of the streams, with the exception of the long, thick Indus and Jhelum streams.
The Indus stream has an offset in $\mu_{\phi_1}$ derived from the by-eye and the GMM measurements, which can be attributed to the significant proper motion gradient fit by the GMM analysis. 
In contrast, the discrepancy in the Jhelum stream can be attributed to the existence of two distinct components of the stream \citep{Bonaca:2019}.
Individual streams are discussed in more detail in \secref{streams}. 

Initial proper motion measurements from the by-eye analysis were used to target \SSSSS (Li et al., submitted).
Seven of the nine streams measured here (Aliqa Uma, ATLAS, Chenab, Elqui, Indus, Phoenix, and Jhelum) have been observed by \SSSSS, and a preliminary analysis of the \SSSSS data shows that the stars used to derive our proper motion measurements have relatively small dispersions in radial velocity space (Li et al. submitted).
We take this as a spectroscopic confirmation of the proper motion measurements quoted here.
An eighth stream, Tucana III, has been previously observed spectroscopically by \citet{Li:2018}, and we again find that the proper-motion members are tightly grouped in radial velocity space.
The ninth stream, Turranburra, has not been fully observed by \SSSSS; however, we find that the proper motion measured here is consistent with the proper motion of RR Lyrae stars observed by \Gaia  that are spatially consistent with the stream.
We describe the analysis of these RR Lyrae in more detail in \secref{streams} and take this to be a secondary confirmation of the proper motion of this stream.

We also report lower-confidence proper motion by-eye measurements for four streams, Ravi, Turbio, Wambelong, and Willka Yaku (\figref{eye_nosignal}). 
Early versions of the by-eye measurements were used to target \SSSSS, but the GMM fits to these streams failed to converge. This suggests that these by-eye measurements are less confident than those mentioned previously.
Upcoming observations from \SSSSS should help resolve the proper motions of these streams.
We find no promising proper motion signal for Molonglo using either method.
The best-fit proper motions and proper motion gradients for all streams are reported in Table \ref{tab:results}, and the by-eye results for all streams, including the low-confidence measurements, are reported in \tabref{results_eye}.

\section{Discussion}
\label{sec:discussion}

\begin{figure*}[t!]
  \centering
    \includegraphics[width=0.8\textwidth]{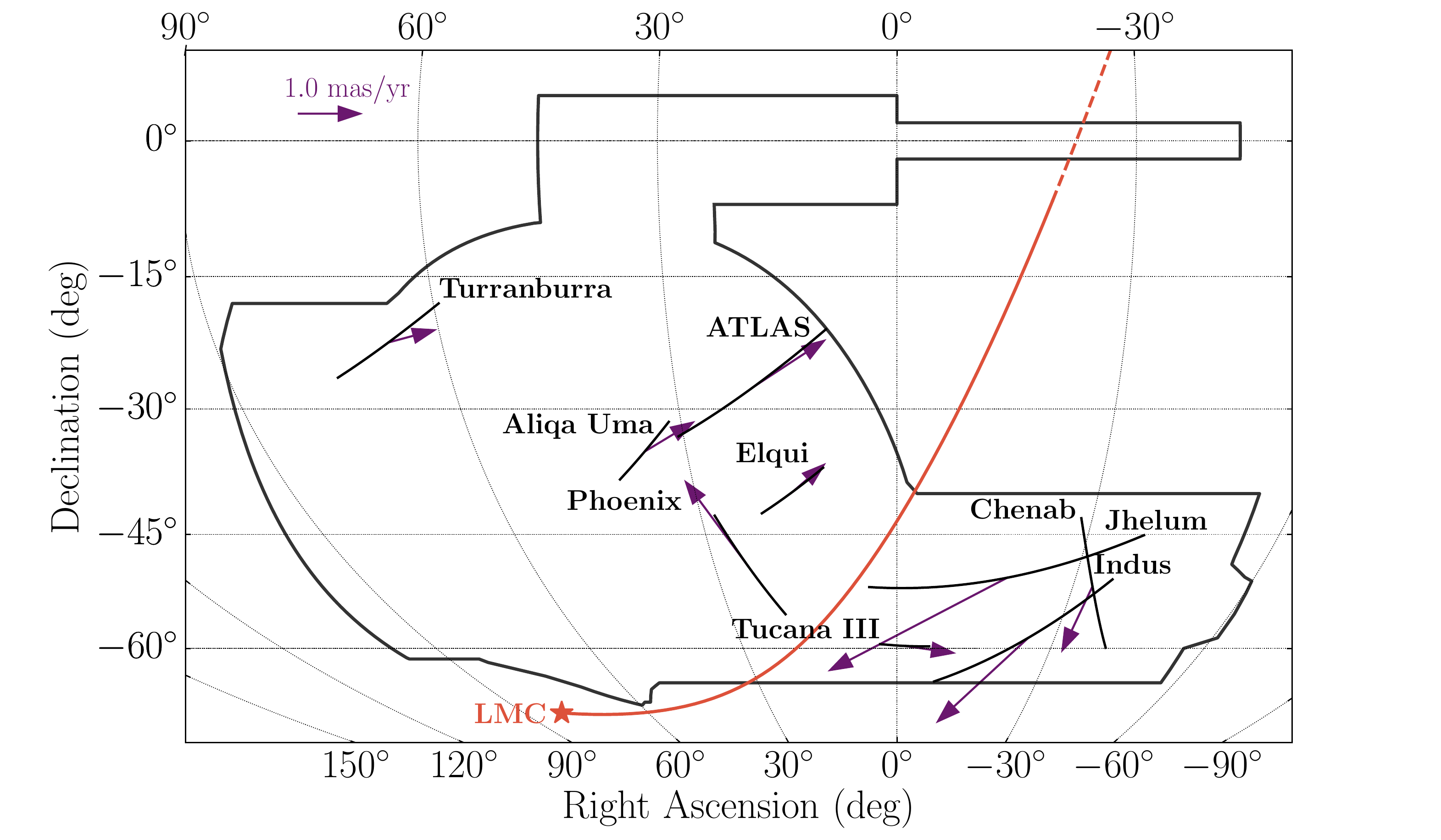}
    \caption{Proper motions offsets in comparison to the orbit of the LMC (orange line). The nine streams with confident proper motion measurements are shown. All but Tucana III and Elqui show significant proper motion offsets. The black lines indicate the stream tracks, as approximated by a great circle. The purple arrows show the reflex-corrected proper motions of the streams. The orange line is the trailing orbit of the LMC over the past 1 Gyr, with the current position marked as an orange star, and the dashed line indicating a distance of greater than 100 kpc. The majority of the proper motion offsets point towards the orbit of the LMC, indicating perturbation by the satellite as a likely cause.}
    \label{fig:lmc}
\end{figure*}

In this section, we discuss the proper motions of the DES streams individually and as a population.
These observations are summarized most concisely in \figref{lmc}, which compares the proper motions measured here to the stream tracks measured with DES imaging. 
The black lines show the stream tracks approximated as great circles passing through the endpoints measured in \citet{Shipp:2018}; the purple arrows show the direction of the proper motion; and the orange line shows the past 1 Gyr of the trailing orbit of the LMC, with the star indicating its present-day position, and the dashed line indicating the segment of the orbit at which the LMC is at a distance beyond 100 kpc. This orbit is performed in the standard \code{MWPotential2014} from \code{galpy} \citep{Bovy:2015} with LMC proper motions from \cite{kallivayalil_etal_2013}, distance from \cite{pietrzynski_lmc_dist}, and radial velocity from \cite{vandermarel_lmc_rv}.
In the following subsections, we present a discussion of individual streams (\secref{streams}) and the influence of the LMC (\secref{lmc}).

\subsection{Discussion of Individual Streams}
\label{sec:streams}

\subsubsection{Aliqa Uma}

Aliqa Uma is a narrow stream that is among the 11 streams discovered in the DES by \citet{Shipp:2018}. Aliqa Uma lies in a complicated region, bordering the southern end of the ATLAS stream and passing near to the Fornax dwarf galaxy in projection. For this reason, it is difficult to select likely members of Aliqa Uma without contamination from these nearby populations. 
Aliqa Uma has the lowest Bayes factor of the nine streams presented here, and in fact the Bayes factor is negative. However, confirmation by early \SSSSS observations merits the inclusion of these measurements among the ``high confidence" list.

\subsubsection{ATLAS}

\begin{figure*}[t!]
  \centering
    \includegraphics[width=0.8\textwidth]{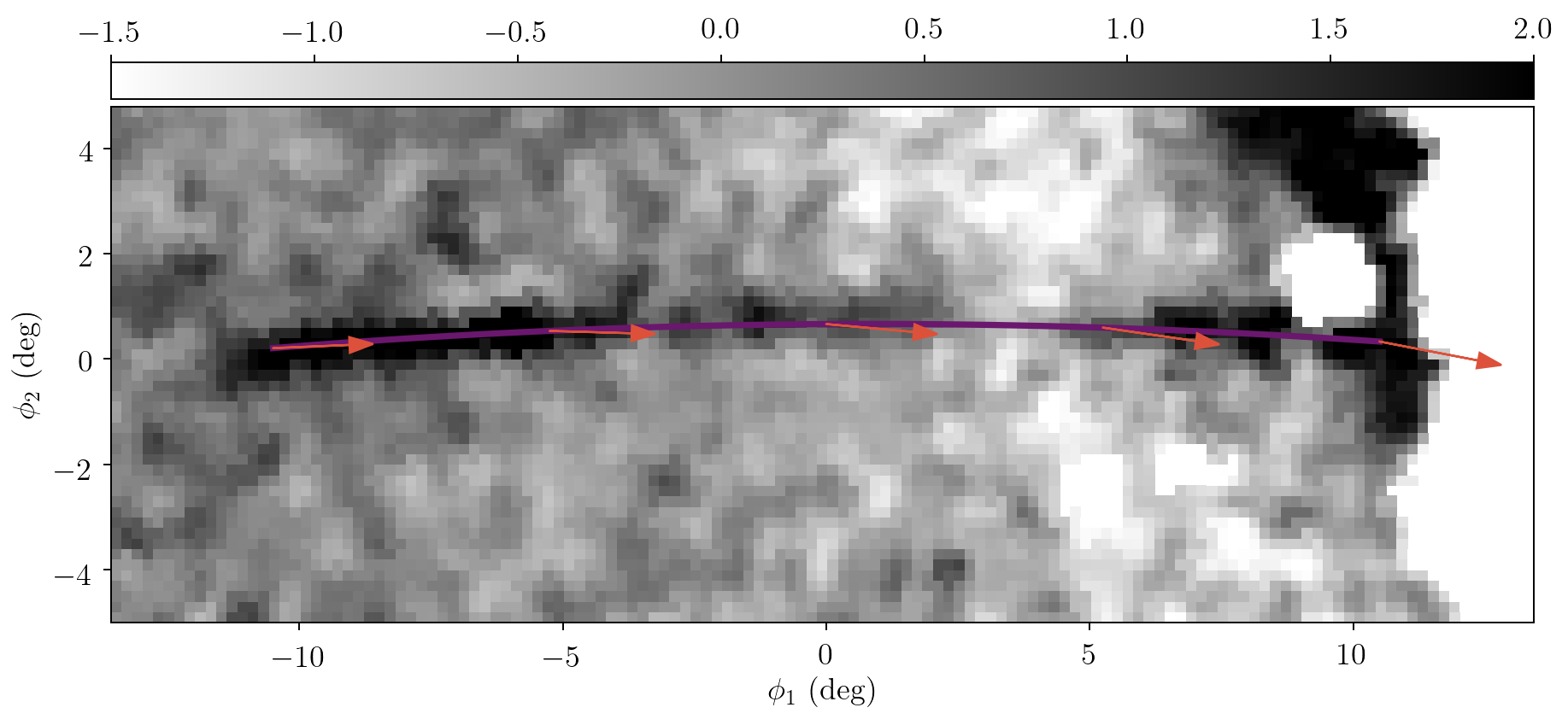}
    \caption{Proper motions offsets along the track of the ATLAS stream. The purple dashed line indicates the track of ATLAS, and the orange arrows show the direction of the proper motion at points along the stream. The offset between the track and proper motion varies along the length of the stream.}
    \label{fig:atlas}
\end{figure*}

The ATLAS stream is the most significant narrow stream in the DES footprint. Originally discovered in data from the ATLAS survey \citep{Koposov:2014}, this stream extends over $>30\degr$ and has been detected by both Pan-STARRS \citep{Bernard:2016} and DES \citep{Shipp:2018}. 
\citet{Shipp:2018} note that the track of the ATLAS stream deviates appreciably from a great circle on the sky. 
Due to the relatively large number of bright stars in ATLAS, it is possible to measure proper motions at multiple positions along the curved track of the stream. 
We note that the offset between the stream track and the proper motion changes along the path of the stream (\figref{atlas}).

The ATLAS and Aliqa Uma streams are nearly adjacent, but are offset by $\roughly 6\degr$ in apparent orbital orientation and $\roughly 6 \kpc$ in mean distance \citep{Shipp:2018}.
However, we find that the reflex-corrected proper motions of these two streams, $\mu'_\alpha \cos\delta, \mu'_\delta = -1.47, 0.78$  mas/yr for ATLAS and $\mu'_\alpha \cos\delta, \mu'_\delta = -0.95, 0.42$  mas/yr for Aliqa Uma, are found to be nearly aligned on the sky (\figref{lmc}).
A potential association has been noted using preliminary radial velocity data from \SSSSS (Li et al., submitted).

\subsubsection{Chenab}

\begin{figure*}[t]
    \centering
    \gridline{\fig{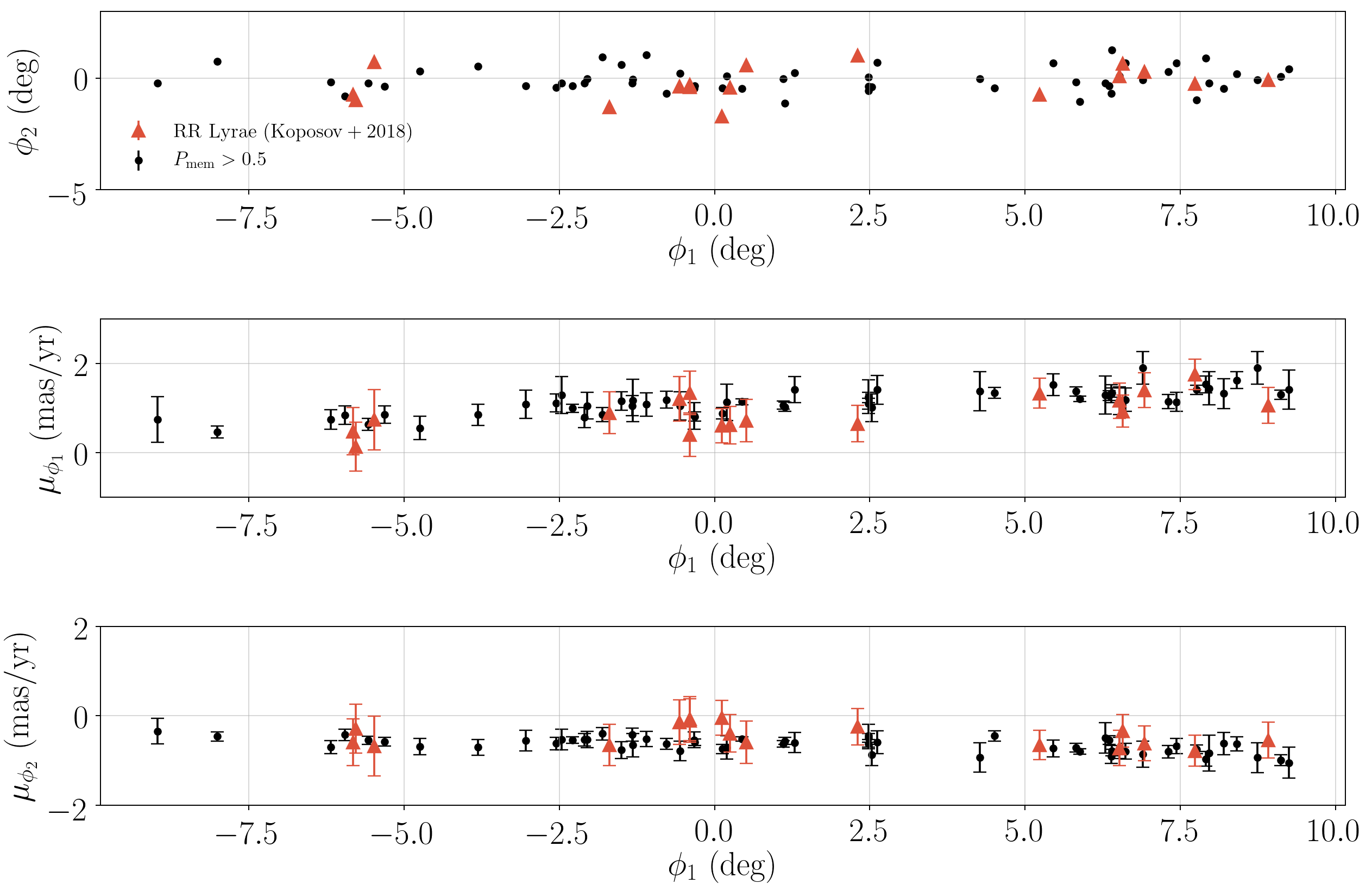}{0.55\textwidth}{}
              \fig{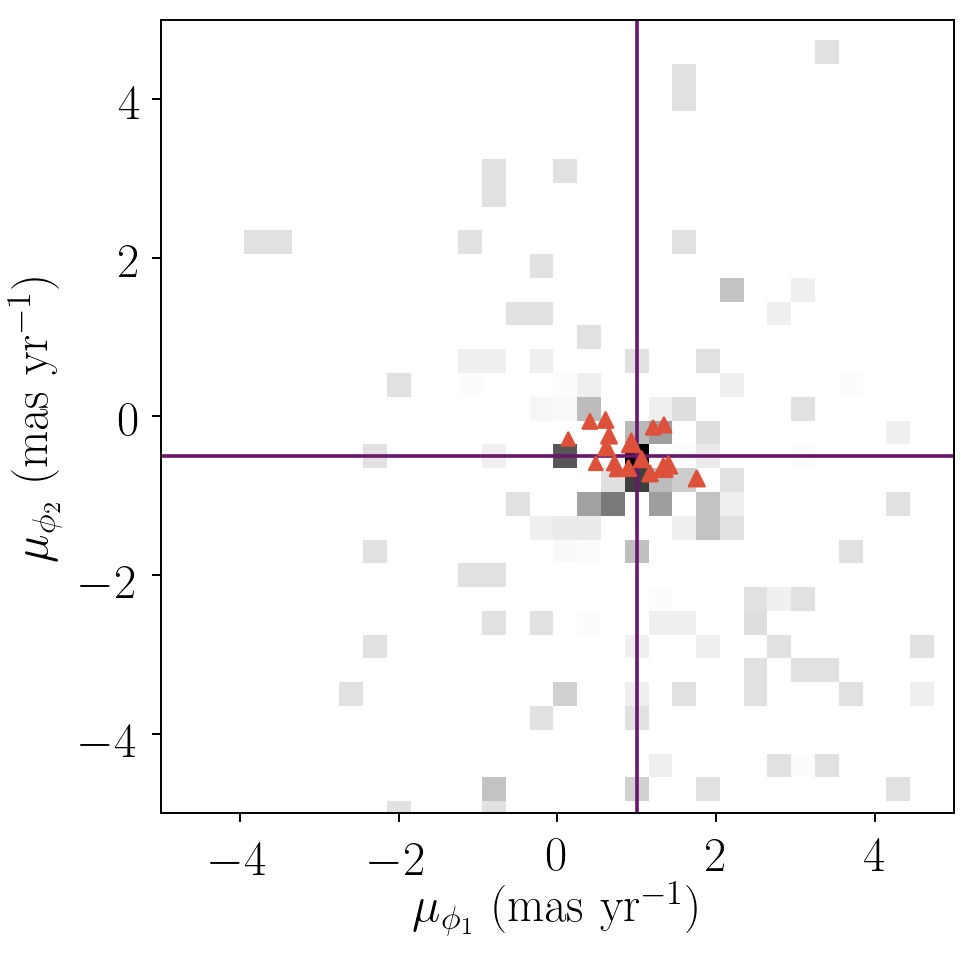}{0.35\textwidth}{}}
    \caption{Comparison between the RGB proper motion measurement of Chenab reported here, and the RR Lyrae members reported by \citet{Koposov:2019}. The black points are stars with $\mathrm{P_{\rm mem}} > 0.5$ from the GMM analysis, and the orange triangles are the RR Lyrae. On the right, the purple crosshair is the GMM proper motion measurement reported here. We find that the high membership probability RGB stars are consistent in proper motion with the reported RR Lyrae members, and generally have smaller proper motion uncertainties.}
    \label{fig:chenab_rrl}
\end{figure*}

The Chenab stream was originally discovered photometrically with data from DES. 
Recently, using measurements of RR Lyrae stars from \Gaia DR2, \citet{Koposov:2019} showed evidence for a Southern Galactic extension of the Orphan stream that overlaps with Chenab.
We independently measure the proper motion of the Chenab stream and find that the proper motions of the RGB stars in Chenab are consistent with those of the RR Lyrae identified by \citet{Koposov:2019}.
We show a comparison between the RGB and RR Lyrae members in \figref{chenab_rrl}. 

The association between Chenab and the Orphan stream was initially unclear due to the $>20\degr$ offset between their Galactocentric orbital poles \citep{Shipp:2018}. 
\citet{Erkal:2018b} showed that this shift in the orbital pole can be caused by the perturbative influence of the LMC. Moreover, this effect is strongest for the southern extension of the Orphan stream, i.e. Chenab, which has passed closer to the LMC than the northern extension. Therefore, these RGB candidate members in Chenab are ideal targets for spectroscopic followup to probe the effect of the LMC in 6D phase space, as the line-of-sight velocities of RR Lyrae are difficult to obtain. 
Furthermore, the large number of bright RGB members improves the precision of the proper motions of Chenab to better constrain the mass of the LMC.

\subsubsection{Elqui}
Elqui is the most distant stream discovered in DES at a distance of $\roughly 50\kpc$.  
\citet{Shipp:2018} suggest that the location, distance, and orientation of Elqui may be a signature of a possible association with the Magellanic Stream. Though the distance is similar to that of the LMC, we find here from its proper motion that Elqui is moving in the opposite sense to the direction of LMC infall (see Figure \ref{fig:lmc}).
This makes it unlikely that Elqui originated as part of the Magellanic System.
It is also unlikely that Elqui would have experienced a temporally extended encounter with the LMC, making it unlikely for the LMC to impart a large gravitational perturbation on the motion of the stream.
This is similar to how streams on retrograde orbits are distorted less by the bar \citep[e.g.][]{Hattori:2016} or substructure in the Milky Way disk \citep[][]{Amorisco:2016} than streams on prograde orbits.  
Indeed, we see here no significant proper motion offset with respect to the stream track: $\mu_{\phi_2} = -0.03 \pm 0.05$ mas/yr.

\subsubsection{Indus}

We find that Indus has the largest measured change in proper motion along the stream, with a gradient of $d \mu_{\phi_1}/d \phi_1, d \mu_{\phi_2}/d \phi_1  = 0.05, 0.04$ mas/yr/deg, resulting in a total change of $1.0$ and $0.8$ mas/yr across the $20\degr$ length of the stream, respectively.

\citet{Malhan:2018b} recover Indus within the \Gaia DR2 data. They find a proper motion range of $0.50 < \mu_{\alpha} \cos{\delta} < 6.0$ mas/yr, $−8.0 < \mu_{\delta} < −2.0$ mas/yr. We find this to be consistent with our by-eye measurement, given uncertainties and the significant proper motion gradient. 
Fitting the GMM to proper motions in the observed frame without correcting for solar reflex motion fails to converge due to the large extent and proper motion gradient of Indus.

\citet{Bonaca:2019} find that the track of the Indus stream is matched to an orbit fit of Jhelum, suggesting that the two streams may be multiple wraps of the same system. The proper motions and their gradients reported here may be used to further explore this possible scenario as discussed below.

We observe an offset between the track of Indus and the direction of its proper motion. However, we note that this offset can be accounted for by a change in distance modulus of 0.2 mag, which is within reasonable uncertainty on the distance modulus measurement obtained by isochrone fitting in \citet{Shipp:2018}.

\subsubsection{Jhelum}

\begin{figure}[t!]
  \centering
    \includegraphics[width=0.9\columnwidth]{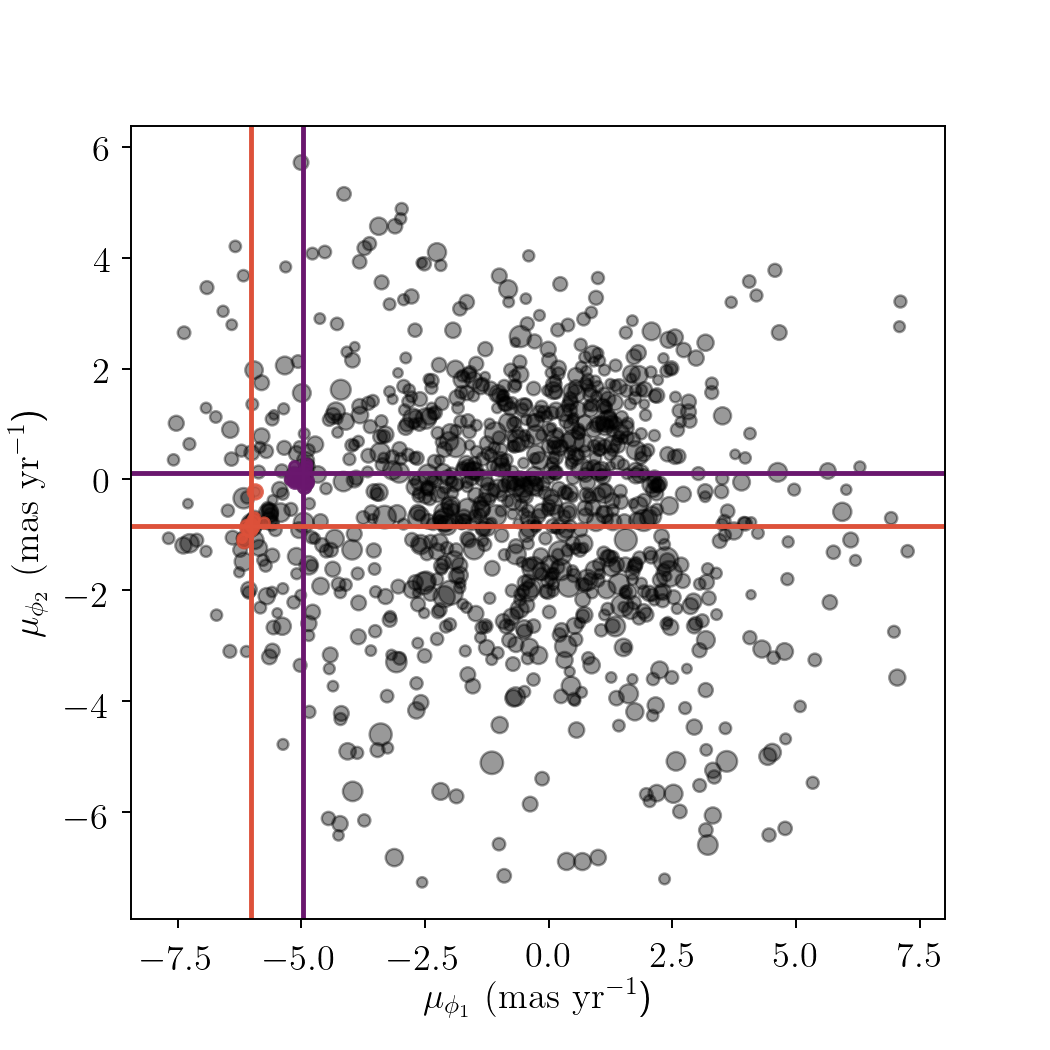}
    \caption{Proper motion of stars around Jhelum. The black points are the stars passing the by-eye cuts described in \secref{data}. The purple and orange crosshairs indicate the best-fit mean proper motions of the two stream components. The purple and orange points are stars with $P_{\rm mem} > 0.5$ for each of the two components.}
    \label{fig:jhelum}
\end{figure}

In on-sky coordinates, Jhelum is the longest ($29.2\degr$) and widest ($1.16\degr$) stellar stream discovered by DES.
We find that the proper motion of Jhelum is best-fit by two distinct components (Figure \ref{fig:jhelum}). 
We fit the two proper motion components simultaneously by introducing a second Gaussian stream component to our model with the same spatial prior as the first component, but with an independent proper motion. We label the two components Jhelum-a and -b in Table \ref{tab:results}. The Bayes factor between the two-stream and one-stream models is 13.5, indicating a significant preference for the two-stream model.
The proper motion of the Jhelum-a component is found to be in good agreement with the by-eye value.
We note that for Jhelum, as for Indus, fitting to the proper motions in the original observed frame without correcting for solar reflex motion fails, and thus these observed frame proper motions are left out of \tabref{results}. The by-eye measurement in the observed coordinate frame is included in \tabref{results_eye}.

We also explored the effect of introducing a distance offset between the two components of the stream. 
Due to the reflex motion correction, there is a degeneracy between the distance separation of the two components and the observed proper motion offset.
We find that the mean proper motions of the two components would converge when corrected for the solar reflex motion at a difference in distance modulus of 0.7 ($m-M = 15.6 \pm 0.35$). 
We note, however, that such discrepant distances ($11\kpc$ and $15\kpc$ for the two components) would require a low probability coincidence in alignment between the two components and the line of sight. 
A smaller distance separation between the two components is possible and could reduce the proper-motion offset slightly.

Interestingly, \citet{Bonaca:2019} recently showed that Jhelum has an extended two-component spatial structure. However, they find the two components to have consistent proper motions. Meanwhile, we measure two distinct proper motion components with consistent spatial distributions.
\citet{Bonaca:2019} has explored possible physical scenarios for the formation of the complex morphology of Jhelum. For additional insight into the physical origin of the two populations, we call attention to the extensive work on the Sagittarius dwarf tidal stream(s) which have been noted to be split into at least two roughly parallel components at slightly different distances in the leading tail in the North \citep{Belokurov:2006}.  A similar split was then noted in the trailing tail in the South by \citet{Koposov:2012}.  \citet{Navarrete_2017} argue that the two southern components are not different wraps of the Sagittarius stream, but could result from either complex or compound structure within the Sagittarius progenitor, or possibly a past interaction with another system, such as the Cetus Polar stream \citep{Newberg:2009}.

The GD-1 stream also has a complex morphology, which may have been caused by past interactions \citep{Carlberg:2013,de_Boer:2018,Price-Whelan:2018}.   By analogy, we note the possibility that Jhelum, too, could be either a) a compound structure (two previously bound objects moving on similar orbits) or b) have been originally a single object, which due to a close interaction with another body, becomes split into two or disrupted or tidally extended so that it now appears like an object with a pair or range of proper motions. Radial velocities and velocity dispersion measurements of the Jhelum components, along with more detailed orbital modeling of Jhelum, as well as comparison with other halo objects, may be able to differentiate between possibilities a) and b).

Interestingly, Jhelum is on a nearly polar orbit with respect to the Milky Way disk \citep{Shipp:2018}. \cite{Erkal:2016} showed that such streams are the most sensitive to the flattening of the halo if the flattening is aligned with the Milky Way disk. This occurs due to differential precession of the stars in the stream and causes the stream to rapidly fan out. Thus, Jhelum's broad morphology in proper motion could be a sensitive probe of the flattening of the Milky Way halo. This will be revisited in future work with radial velocities from \SSSSS (Li et al., submitted). 

We also consider the possible effect of the Indus stream on Jhelum. The distance modulus of Indus is 16.1, while that of Jhelum is 15.6 \citep{Shipp:2018}, a 50\% difference in distance, and both are traveling in roughly the same direction (see Figure \ref{fig:lmc}). 
Therefore one may consider whether these two streams could have had a close encounter in the past or perhaps share a common origin. \citet{Bonaca:2019} find that an orbit fit to one component of Jhelum passes through the track of the Indus stream, which may indicate that the two streams are different tidal debris wraps from a common progenitor, or that a close approach has occurred between two distinct streams. A close encounter could explain the double structure of Jhelum, though one would require additional radial velocity information from both systems to more confidently determine their orbital histories.

\subsubsection{Phoenix}

The Phoenix stream was first discovered by \citet{Balbinot:2016} using data from the first year of DES.
Compared to the other DES streams, the stellar distribution of Phoenix appears considerably more clumpy; however, none of these overdensities has been conclusively determined to be associated with a Phoenix progenitor. 
We examine the \Gaia data for evidence of the structures identified by \citet{Balbinot:2016}; however, the \Gaia stars passing our selections are too sparse to resolve any of these excesses.

\citet{Grillmair:2016b} speculated that the Hermus stream \citep{Grillmair:2014} may be a northern extension of Phoenix.
They predict that if Phoenix-Hermus were one stream on a prograde orbit, it would have a proper motion of $\mu_{\alpha}\cos{\delta},\mu_{\delta} \sim 2.1, 0.1$ mas/yr, while a retrograde orbit would yield $\mu_{\alpha}\cos{\delta},\mu_\delta \sim 1, -3.5$ mas/yr.
Our measured proper motion of $\mu_{\alpha}\cos{\delta},\mu_\delta \sim 2.76 \pm 0.02, -0.05 \pm 0.02$ mas/yr disfavors the retrograde model. 

\citet{Balbinot:2016} also note a possible association between Phoenix and the nearby globular cluster, NGC 1261.
Using proper motion measurements from \citet{Dambis:2006}, $\mu_{\alpha} \cos{\delta}, \mu_\delta = 1.33 \pm 0.89, -3.06 \pm 1.06$ mas/yr, \citet{Balbinot:2016} find that NGC 1261 is on an orbit aligned with the path of Phoenix, but offset by $\roughly 10\degr$.
Recently, \citet{Vasiliev:2019} used \Gaia DR2 to update the proper motion of NGC 1261, yielding a value of $\mu_{\alpha} \cos{\delta}, \mu_\delta = 1.632 \pm 0.057, -2.037 \pm 0.057$ mas/yr (consistent values were determined by \citealt{Baumgardt:2019}).
The combination of this updated proper motion measurement for NGC 1261 and our measurement of the proper motion of the Phoenix stream make it increasingly unlikely that these two systems share a physical origin.
However, we do note that the proper motion offset of Phoenix is slightly aligned towards the orbit of NGC 1261.

\subsubsection{Tucana III}

The Tucana III stream is composed of two tidal tails extending from the Tucana III dwarf galaxy \citep{Drlica-Wagner:2015}, and is the only stream in the DES footprint with a definitive progenitor.
The proper motion of the Tucana III dwarf galaxy has been measured by several groups: 
\citet{Pace:2018} find $\mu_{\alpha} \cos{\delta}, \mu_{\delta} = -0.03 \pm 0.04, -1.65 \pm 0.04$ mas/yr, \citet{Simon:2018} finds $\mu_{\alpha} \cos{\delta}, \mu_{\delta} = -0.014 \pm 0.038, -1.673 \pm 0.040$ mas/yr, and \citet{Fritz:2018} find $\mu_{\alpha} \cos{\delta}, \mu_{\delta} = −0.025 \pm 0.034 \pm 0.035, −1.661 \pm 0.035 \pm 0.035$ mas/yr.
The Tucana III stream is expected to have a similar proper motion to the dwarf itself. 
We measure a proper motion for Tucana III, including the core and tidal tails, of $\mu_{\alpha} \cos{\delta}, \mu_{\delta} = -0.10 \pm 0.04,  -1.64 \pm 0.04$ mas/yr, which is indeed similar to that of the Tucana III dwarf galaxy. We also find that Tucana III has the largest proper motion gradient of the streams measured here, with $d \mu_{\phi_1}/d \phi_1, d \mu_{\phi_2}/d \phi_1 = 0.12 \pm 0.03, -0.06 \pm 0.03$ mas/yr/deg.
 
\citet{Erkal:2018a} fit the orbit of Tucana III based on the track of the Tucana III stream and the line-of-sight velocity from \citet{Li:2018}, prior to \Gaia DR2. 
They argued that the orbit of Tucana III was likely perturbed by a recent close passage with the LMC. 
They predicted that the LMC would have induced a non-zero proper motion perpendicular to the track of the stream and that this non-zero proper motion could be used to constrain the mass of LMC. 
However, our measurements show that the proper motion perpendicular to the stream, $\mu_{\phi_2} = -0.03 \pm 0.03$ mas/yr, is consistent with zero. 
Since the lack of a proper motion perpendicular to the Tucana III stream track would set an upper bound on the mass of the LMC that is inconsistent with other direct measurements \citep[e.g.][]{vanderMarel:2014}, we suggest three possible explanations for the discrepancy between our measurements and the model of \citet{Erkal:2018a}. 
First, $\mu_{\phi_2}$ is corrected for the solar reflex motion, and is therefore distance-dependent. 
The apparent lack of a perpendicular proper motion might indicate that Tucana III is more distant than the initial isochrone fits suggest. In fact, a similar suggestion was made by \citet{Erkal:2018a} based on preliminary measurements of 4 RR Lyrae stars in Tucana III.
Second, the lack of offset may be due to the fact that \citet{Erkal:2018a} fit the orbit with a fixed Milky Way potential. 
The proper motions will also be sensitive to the potential and the mass of the Milky Way. 
Third, \citet{Erkal:2018a} did not consider the reflex motion of the Milky Way caused by the infall of the LMC. 
As shown in \citet{Erkal:2018b}, the distance and speed of the Milky Way relative to its present day position and velocity is non-negligible, which will affect the modeling of the proper motion of Tucana III.

\subsubsection{Turranburra}
\label{sec:turranburra}

Turranburra is a relatively thick stream located at the eastern edge of the DES footprint. 
The morphology of the stream suggests a dwarf galaxy progenitor;  \citet{Shipp:2018} predict a progenitor mass of $1.8 \times 10^{6}\ \mathrm{M_{\odot}}$.
Interestingly, in spite of its distance from the LMC, Turranburra also shows an appreciable offset between its track and observed proper motion, which is directed towards the LMC.

\begin{figure*}
    \centering
    \gridline{\fig{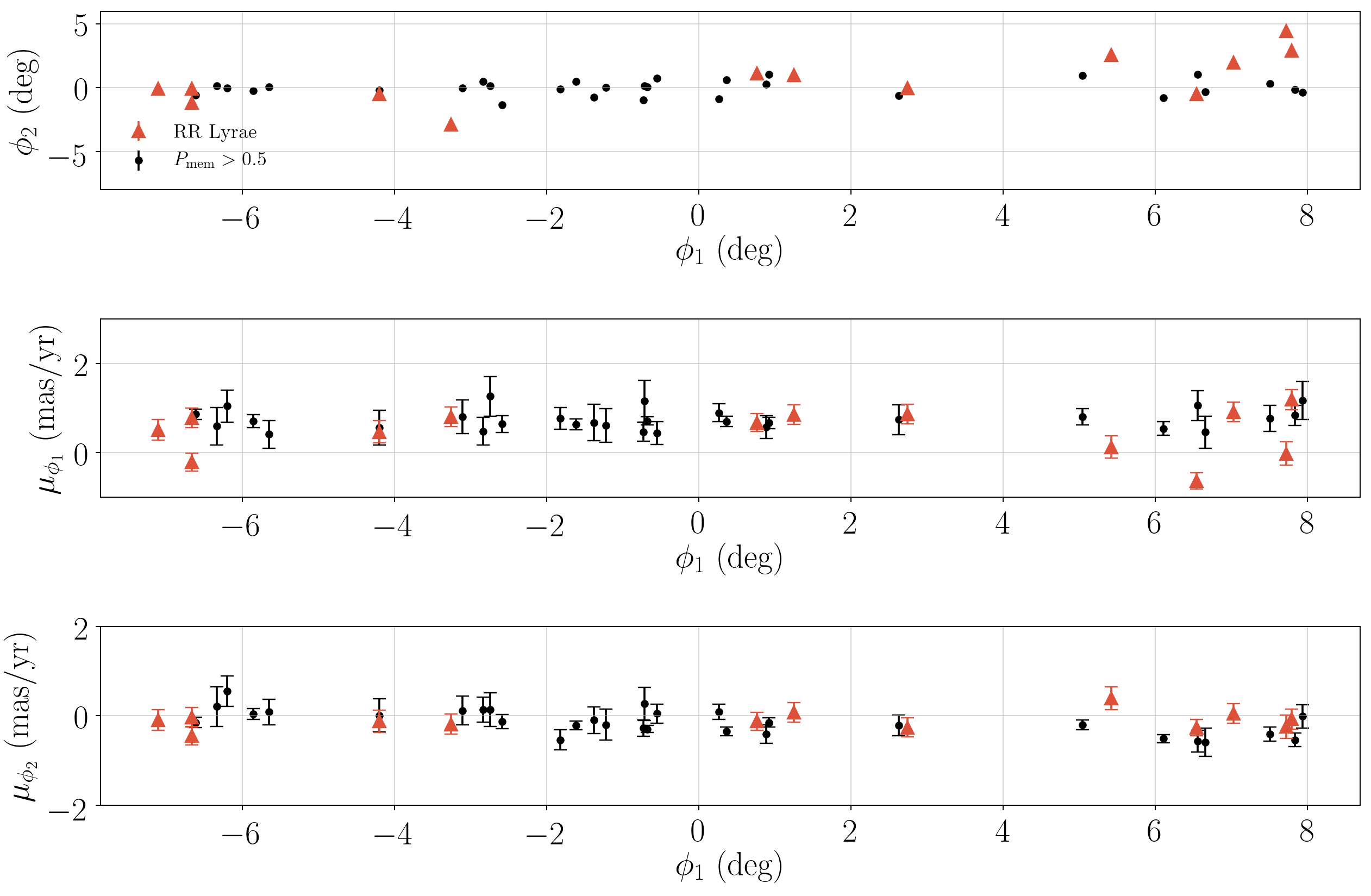}{0.55\textwidth}{}
              \fig{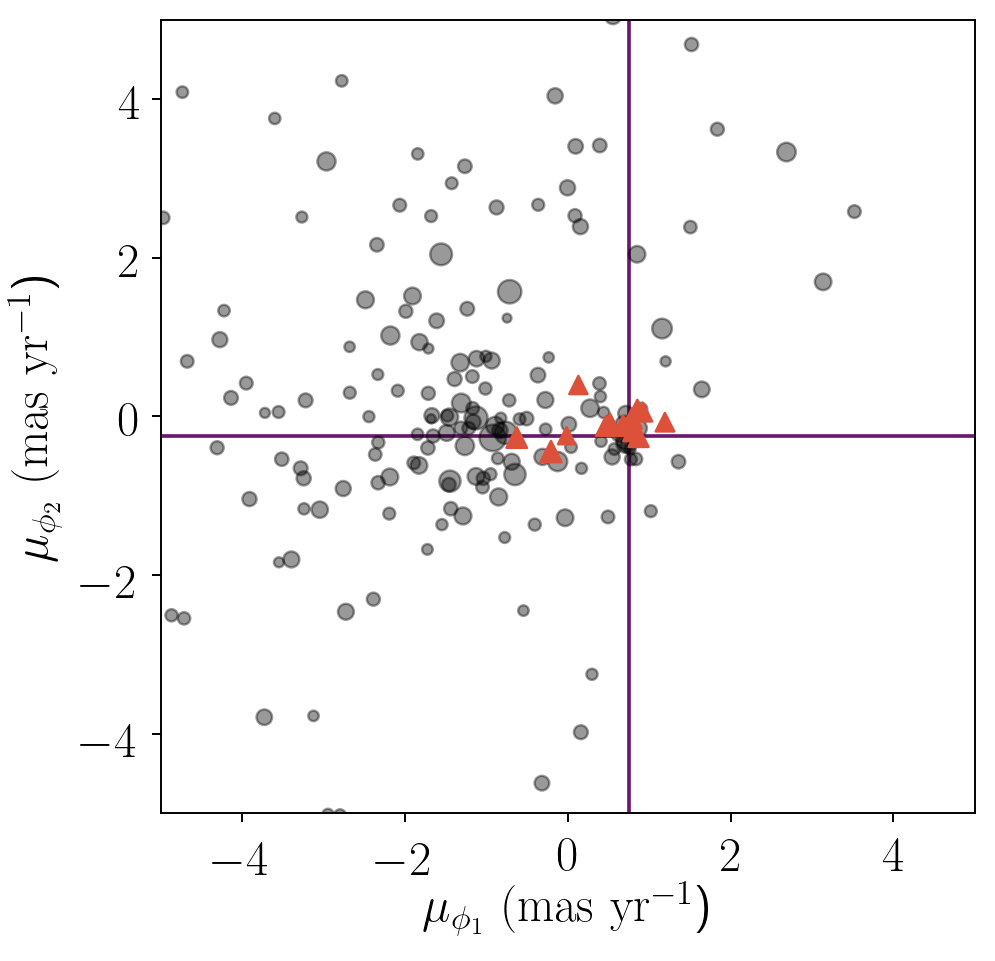}{0.35\textwidth}{}}
    \caption{Comparison between the RGB proper motion measurement of Turranburra, and the RR Lyrae members selected as described in \secref{turranburra}. The black points are stars with $\mathrm{P_{\rm mem}} > 0.5$ from the GMM analysis, and the orange triangles are the RR Lyrae. On the right, the purple crosshair is the GMM proper motion measurement reported here. We find that the high membership probability RGB stars are consistent in proper motion with the selected RR Lyrae.}
    \label{fig:turranburra_rrl}
\end{figure*}

Unlike the other eight streams previously mentioned, Turranburra has not yet been fully observed by \SSSSS and we cannot confirm its proper motion signature with radial velocities. 
However, we have independently confirmed the proper motion measurement by comparison to the sample of \Gaia DR2 RR Lyrae published by \citet{Iorio:2018}. 
We find 12 RR Lyrae that are likely associated with the stellar stream (see Figure \ref{fig:turranburra_rrl}).

The RR Lyrae were selected first along the length of the stream ($\vert \phi_{\rm 1} \vert < 8.5\degr$) and within $\vert \phi_{\rm 2} \vert < 5\degr$. Then, we selected stars within $3 \kpc$ of the distance to Turranburra reported in \citet{Shipp:2018}. The distances to the RR Lyrae are calculated using Equation 2 in \citet{Iorio:2018}.
The RR Lyrae passing these simple selections, which are listed in Appendix \ref{app:turranburra_rrl}, all lie within $2.5\degr$ of the stream track, and the majority are tightly clustered around the measured value of the proper motion of the stellar stream.

\subsubsection{Ravi}

While the Ravi stream was not detected with high confidence in our analysis, we do note an interesting association with the RR Lyrae stream 24.5-1 from \citet{Mateu:2018}.
The close association in the orbital poles of these two streams was previously noted by \citet{Shipp:2018}.
We take the RR Lyrae stars associated with 24.5-1 as reported by \citet{Mateu:2018}, and select stars that lie within $50\degr$ along the stream track of the mid-point of Ravi. We determine the median proper motion of these RR Lyrae to be $\mu_{\rm \alpha} \cos{\delta}, \mu_{\delta} \sim 0.6, -1.8$ mas/yr, with a large spread in $\mu_{\rm \alpha} \cos{\delta}$ of $\sim 0.5$ mas/yr. This value is similar to our low-confidence proper motion measurement for Ravi of $\mu_{\rm \alpha} \cos{\delta}, \mu_{\rm \delta} \sim 0.2, -1.6$ mas/yr, particularly given the imprecision of the by-eye measurement, which may be further indication of an association between these two streams.

We also note that Ravi crosses the dwarf galaxy Tucana II \citep{Bechtol:2015, Koposov:2015} in projection, although the galaxy is at a much larger distance. (Tucana II is at a distance of 57 kpc \citep{Koposov:2015}, while Ravi is at a distance of 23~kpc.)
In order to exclude contamination from Tucana II, we selected only the segment of Ravi with $\phi_1 > 0\degr$, so that the closest stars to Tucana II are separated from the galaxy by $> 7\degr$. However, the proper motion we measure for Ravi is similar to that of Tucana II, $\mu_{\rm \alpha} \cos{\delta}, \mu_{\delta} \sim 0.91, -1.16$ mas/yr \citep{Pace:2018}, which could indicate that Tucana II has a very extended stellar distribution that is contaminating our analysis, or that the two systems share similar proper motions despite their large physical separation.

\subsubsection{Other Streams}
We do not find high-confidence measurements for three other streams: Turbio, Wambelong, and Willka Yaku, and we find no good measurement for Molonglo.  
We note that these streams reside in complex regions of higher stellar density, either nearer to the Galactic plane, or in areas with other known streams or halo structure.
Wambelong is located at $b \sim -30\degr$, while Turbio and Willka Yaku are both in the vicinity of the Eri-Phe overdensity \citep{Li:2016}. 
Molonglo is both near to the ATLAS stream and stretches into the area covered by the complex, massive Sagittarius stream.
The complexity of the stellar foreground in these regions may have contributed to the failure of the the GMM fit to converge on valid proper motion measurements for these streams.

\subsection{Influence of the LMC}
\label{sec:lmc}

The LMC is the largest satellite of the Milky Way and can significantly perturb the orbits of stellar streams \citep{Erkal:2018b}.
One possible consequence of a perturbation by the LMC is a misalignment between a stream’s track on the sky and the direction of its motion.
\citet{Erkal:2018b} showed that the observed proper motion of the Orphan stream could be explained by a large gravitational perturbation from the LMC.
Observation of the Orphan stream can thus be used to constrain the total mass of the LMC to be $1.38^{+0.27}_{-0.24} \times 10^{11} \Msun$ \citep{Erkal:2018b}. 
\citet{Erkal:2018a} hypothesized that such a massive LMC would similarly perturb the Tucana III stream; however, such a perturbation has not been found. 
The extent of the LMC perturbations on other streams is still unknown, and detailed modeling will be required to develop a self-consistent scenario.
However, we can use the observed stream tracks and proper motions to make a qualitative comparison.

In \figref{lmc}, the majority of the streams show proper motion offsets in the direction of the LMC, with the exception of Phoenix, whose offset is pointing away from the LMC, and Elqui and Tucana III, which show no significant offsets. However, we note that we are only considering two dimensions of the stream velocity; it is also possible for these streams to have experienced a perturbation to their radial velocities, which will be explored in more detail by \SSSSS.

The magnitude of the measured proper motion offset for a stream depends on the distance assumed in the solar reflex correction. We find that in addition to Elqui and Tucana III (which show very small proper motion offsets), only Indus has an offset that is consistent with zero given a characteristic uncertainty in distance modulus of 0.2 mag. The other streams would require changes in distance modulus ranging from $\roughly 0.5 - 4$ mag to account for the observed proper motion offsets.

A visual inspection of \figref{lmc} and \tabref{results} gives anecdotal evidence that certain groups of streams may exhibit more significant deflections. 
Streams with right ascension  west of the LMC seem to exhibit larger offsets than those to the east of the LMC. Streams with larger widths, which may be indicative of a dwarf galaxy progenitor, generally seem to have larger offsets as well. In addition, streams with proper motion vectors roughly aligned with the direction of motion along the trailing orbit of the LMC also exhibit larger offsets. These possible relationships must be examined in more detail with radial velocities and detailed modeling.
 In any case, the offsets seen here indicate that the DES streams are excellent candidates for placing strong constraints on the LMC mass, as well as its shape and radial density profile.

Offsets between the track and direction of motion of stellar streams can also be caused by time-dependent oscillations in the Milky Way's potential due to recent accretion events, as shown by \citet{Carlberg:2019}. The detailed modeling of stellar streams, which will be possible with the combination of these measurements and radial velocities from \SSSSS, will allow for the separation between the effects of the LMC, which may dominate in the southern sky, and other large-scale time-dependent variations in the Milky Way's potential.

\section{Conclusions}
\label{sec:conclusion}

We present confident measurements of the proper motions of nine stellar streams in the DES footprint.
These measurements confirm that these streams are coherent systems and illustrate the combined power of \Gaia DR2 and DES DR1 to measure the velocities of distant, low-surface-brightness streams (out to $\roughly 50 \kpc$). 
In addition, we have obtained low-confidence measurements of proper motions of four additional streams in the DES footprint. 
Further velocity measurements, both of proper motions and radial velocities, are necessary to confirm the remaining population of stellar streams discovered in DES and other photometric surveys.

Many of these streams are observed to have significant offsets between the direction of their tracks on the sky and the direction of their proper motions.
This observation may indicate that the LMC may have significantly perturbed the orbits of these streams, and suggests that this population of streams may be used to place strong constraints on the mass and the radial profile of the Milky Way's largest satellite.
Complete orbit modeling requires full 6D phase-space measurements of each stream. 
The proper motion measurements in this work have been used to efficiently select targets for the ongoing \SSSSS spectroscopic survey, which aims to obtain radial velocities and metallicities of 20 streams in the Southern Hemisphere (Li et al. submitted).

In the future, imaging surveys such as LSST \citep{LSST:2009} and WFIRST \citep{WFIRST:2013} will provide sensitive measurements of fainter and more distant streams.
Wide-area spectroscopic surveys, such as DESI \citep{DESI:2016}, WEAVE \citep{WEAVE:2016}, 4MOST \citep{4MOST:2019}, and/or MSE \citep{MSE:2019} will provide complementary radial velocity measurements.
With complete 6D phase space measurements of large populations of stellar streams, it will be possible to place strong constraints on the distribution of mass in our Galaxy, ranging from low-mass subhalos to the total mass of the Milky Way.

\section{Acknowledgments}

We thank Ana Bonaca, Giuliano Iorio, Cecilia Mateu, and Adrian Price-Whelan for helpful conversations.

This material is based upon work supported by the U.S. Department of Energy, Office of Science, 
Office of Workforce Development for Teachers and Scientists, Office of Science Graduate Student Research 
(SCGSR) program. The SCGSR program is administered by the Oak Ridge Institute for Science and Education 
(ORISE) for the DOE. ORISE is managed by ORAU under contract number DE-SC0014664.

This project was developed in part at the 2019 Santa Barbara Gaia Sprint, hosted by the Kavli Institute for Theoretical Physics at the University of California, Santa Barbara.

This research was supported in part at KITP by the Heising-Simons Foundation and the National Science Foundation under Grant No. NSF PHY-1748958.

NS thanks the LSSTC Data Science Fellowship Program, her time as a Fellow has benefited this work.
TSL is supported by NASA through Hubble Fellowship grant HF2-51439.001 awarded by the Space Telescope Science Institute, which is operated by the Association of Universities for Research in Astronomy, Inc., for NASA, under contract NAS5-26555.
ABP acknowledges generous support from the George P. and Cynthia Woods Institute for Fundamental Physics and Astronomy at Texas A\&M University.
SK is partially supported by NSF grant AST-1813881 and Heising-Simon’s foundation grant 2018-1030.

This work presents results from the European Space Agency (ESA) space
mission Gaia. Gaia data are being processed by the Gaia Data
Processing and Analysis Consortium (DPAC). Funding for the DPAC is
provided by national institutions, in particular the institutions
participating in the Gaia MultiLateral Agreement (MLA). The Gaia
mission website is \url{https://www.cosmos.esa.int/gaia}. The Gaia archive website is \url{https://archives.esac.esa.int/gaia}.

This project used public archival data from the Dark Energy Survey
(DES). Funding for the DES Projects has been provided by the
U.S. Department of Energy, the U.S. National Science Foundation, the
Ministry of Science and Education of Spain, the Science and Technology
Facilities Council of the United Kingdom, the Higher Education Funding
Council for England, the National Center for Supercomputing
Applications at the University of Illinois at Urbana-Champaign, the
Kavli Institute of Cosmological Physics at the University of Chicago,
the Center for Cosmology and Astro-Particle Physics at the Ohio State
University, the Mitchell Institute for Fundamental Physics and
Astronomy at Texas A\&M University, Financiadora de Estudos e
Projetos, Funda{\c c}{\~a}o Carlos Chagas Filho de Amparo {\`a}
Pesquisa do Estado do Rio de Janeiro, Conselho Nacional de
Desenvolvimento Cient{\'i}fico e Tecnol{\'o}gico and the
Minist{\'e}rio da Ci{\^e}ncia, Tecnologia e Inova{\c c}{\~a}o, the
Deutsche Forschungsgemeinschaft, and the Collaborating Institutions in
the Dark Energy Survey.  The Collaborating Institutions are Argonne
National Laboratory, the University of California at Santa Cruz, the
University of Cambridge, Centro de Investigaciones Energ{\'e}ticas,
Medioambientales y Tecnol{\'o}gicas-Madrid, the University of Chicago,
University College London, the DES-Brazil Consortium, the University
of Edinburgh, the Eidgen{\"o}ssische Technische Hochschule (ETH)
Z{\"u}rich, Fermi National Accelerator Laboratory, the University of
Illinois at Urbana-Champaign, the Institut de Ci{\`e}ncies de l'Espai
(IEEC/CSIC), the Institut de F{\'i}sica d'Altes Energies, Lawrence
Berkeley National Laboratory, the Ludwig-Maximilians Universit{\"a}t
M{\"u}nchen and the associated Excellence Cluster Universe, the
University of Michigan, the National Optical Astronomy Observatory,
the University of Nottingham, The Ohio State University, the OzDES
Membership Consortium, the University of Pennsylvania, the University
of Portsmouth, SLAC National Accelerator Laboratory, Stanford
University, the University of Sussex, and Texas A\&M University.
Based in part on observations at Cerro Tololo Inter-American
Observatory, National Optical Astronomy Observatory, which is operated
by the Association of Universities for Research in Astronomy (AURA)
under a cooperative agreement with the National Science Foundation.

This manuscript has been authored by Fermi Research Alliance, LLC under Contract No. DE-AC02-07CH11359 with the U.S. Department of Energy, Office of Science, Office of High Energy Physics. The United States Government retains and the publisher, by accepting the article for publication, acknowledges that the United States Government retains a non-exclusive, paid-up, irrevocable, world-wide license to publish or reproduce the published form of this manuscript, or allow others to do so, for United States Government purposes.

\appendix
\numberwithin{figure}{section}
\numberwithin{table}{section}

\section{By-Eye Results}
\label{app:eye}

Here we report the by-eye measurements for all streams, including those with low-confidence measurements (\tabref{results_eye}). For the nine streams with confident measurements these values are consistent, considering the imprecision of the by-eye measurements, with the GMM results.

\newcommand{\eyecaption}{By-eye results.}
\begin{deluxetable}{l ccccc }[hb!]
\tablecolumns{13}
\tablewidth{0pt}
\tabletypesize{\scriptsize}
\tablecaption{ \eyecaption \label{tab:results_eye}}
\tablehead{ & $\mu_{\rm \alpha} cos \delta$ & $\mu_{\rm \delta}$ & $\mu_{\rm \phi_1}$ & $\mu_{\rm \phi_2}$ \\
 & (mas/yr) & (mas/yr) & (mas/yr) & (mas/yr)}
\startdata
Aliqa Uma       &  0.3 & -0.6 &  1.0 & -0.2 \\
ATLAS           & -0.1 & -1.0 &  1.6 & -0.2 \\
Chenab          &  0.3 & -2.4 &  1.0 & -0.5 \\
Elqui           &  0.1 & -0.4 &  0.6 &  0.0 \\
Indus           &  3.5 & -5.4 & -3.8 &  0.1 \\
Jhelum          &  6.9 & -5.8 & -6.0 & -0.8 \\
Phoenix         &  2.8 & -0.1 & -1.9 & -0.4 \\
Tucana III      & -0.1 & -1.7 &  1.1 &  0.0 \\
Turranburra     &  0.4 & -0.9 &  0.8 & -0.3 \\
[+0.5em]\tableline\\[-1em]
Ravi            &  0.2 & -1.6 & 0.5 &  -0.1 \\
Turbio          &  2.3 &  2.0 & -3.8 & -0.3 \\
Wambelong       &  2.0 & -0.1 & -0.8 & -0.9 \\
Willka Yaku     &  1.1 &  0.3 &  -0.9 &  0.0 \\
Molonglo        & \ldots &  \ldots & \ldots &  \ldots \\
\enddata
\end{deluxetable}

\section{Turranburra RR Lyrae}
\label{app:turranburra_rrl}

Table \ref{tab:turranburra_rrl} lists the Gaia DR2 Source ID's of possible RR Lyrae members of Turranburra. The selection of these RR Lyrae is described in Section \ref{sec:turranburra}.

\newcommand{\trrlcaption}{Turranburra RR Lyrae.}
\begin{deluxetable}{l ccc }[hb!]
\tablecolumns{13}
\tablewidth{0pt}
\tabletypesize{\scriptsize}
\tablecaption{ \trrlcaption
\label{tab:turranburra_rrl}}
\tablehead{\Gaia Source ID & $\alpha$ & $\delta$ \\ & (deg) & (deg)}
\startdata
5091448747454278656 & 64.45486 & -21.02018 \\ 
4881423811590801536 & 72.81505 & -26.62611 \\ 
4894078026492980480 & 70.76238 & -24.85804 \\ 
4881772670311841920 & 73.37994 & -25.64553 \\ 
5097830652242359936 & 61.52814 & -15.87997 \\ 
5094366743938630016 & 60.72077 & -19.45971 \\ 
5097133875404904320 & 61.67817 & -17.08474 \\ 
4881586985989030272 & 73.82559 & -25.84417 \\ 
4899710545386636160 & 66.38172 & -20.90392 \\ 
3176477345911441024 & 62.41282 & -14.63724 \\ 
5096494402017554816 & 63.44298 & -17.42091 \\ 
4899649801666240896 & 66.89291 & -21.04826 \\ 
4891992802690351232 & 68.56002 & -26.50784 \\ 
\enddata
\end{deluxetable}

\section{Selection Parameters}
\label{app:sel}

\tabref{isochrones} gives the isochrone parameters used in the data selections described in \secref{data}. These were modified from the parameters reported in \citet{Shipp:2018}, based on visual comparison of high-probability members after a first iteration of the proper motion fit. 

\newcommand{\isocaption}{Isochrone parameters.}
\begin{deluxetable}{l cccc }[h!]
\tablecolumns{13}
\tablewidth{0pt}
\tabletypesize{\scriptsize}
\tablecaption{ \isocaption \label{tab:isochrones}}
\tablehead{Name & $m-M$ & Age (Gyr) & Z}
\startdata
Aliqa Uma       & 17.3 & 12.5 & 0.0001 \\
ATLAS           & 16.8 & 12.5 & 0.0001 \\
Chenab          & 18.0 & 12.5 & 0.0001 \\
Elqui           & 18.5 & 12.5 & 0.0001 \\
Indus           & 16.1 & 12.5 & 0.0004 \\
Jhelum          & 15.6 & 12.5 & 0.0001 \\
Molonglo        & 16.8 & 13.5 & 0.001  \\
Phoenix         & 16.4 & 12.5 & 0.0001 \\
Ravi            & 16.8 & 13.5 & 0.0003 \\
Tucana III      & 17.0 & 13.5 & 0.0001 \\
Turbio          & 16.1 & 12.5 & 0.0001 \\
Turranburra     & 17.2 & 13.5 & 0.0003 \\
Wambelong       & 15.9 & 11.0 & 0.0001 \\
Willka Yaku     & 17.7 & 11.0 & 0.0006 \\
\enddata

\end{deluxetable}

\section{Coordinate Transformation Matrices}
\label{app:matrix}

\newcommand{\matrixcaption}{Rotation matrix parameters.}

We used $3 \times 3$ rotation matrices to transform positions and proper motions from celestial coordinates to stream coordinates. The entries of these matrices, $R_{i,j}$, are shown in Table \ref{tab:matrix}.

The matrices are written as, 

\begin{equation}
R=
\begin{bmatrix}
R_{0,0} & R_{0,1} & R_{0,2} \\
R_{1,0} & R_{1,1} & R_{1,2} \\
R_{2,0} & R_{2,1} & R_{2,2} \\
\end{bmatrix} 
\, . \\ 
\end{equation}

\newcommand{\matrixcomments}
{All transformations are defined by the stream endpoints reported by \citet{Shipp:2018}, with the origin located at the center of the stream, apart from that of Tucana III, for which we use the matrix from \citet{Li:2018}, which centers the stream on the progenitor.} 
\begin{deluxetable}{l ccccccccc}[h!]
\tablecolumns{13}
\tablewidth{0pt}
\tabletypesize{\scriptsize}
\tablecaption{ \matrixcaption }
\tablehead{Name & $R_{0,0}$ & $R_{0,1}$ & $R_{0,2}$ & $R_{1,0}$ & $R_{1,1}$ & $R_{1,2}$ & $R_{2,0}$ & $R_{2,1}$ & $R_{2,2}$}
\startdata
Aliqa Uma       &  0.66315359 &  0.48119409 & -0.57330582 &  0.74585903 & -0.36075668 &  0.55995440 & -0.06262284 &  0.79894109 &  0.59814004 \\
ATLAS           &  0.83697865 &  0.29481904 & -0.46102980 &  0.51616778 & -0.70514011 &  0.48615660 &  0.18176238 &  0.64487142 &  0.74236331 \\
Chenab          &  0.51883185 & -0.34132444 & -0.78378003 & -0.81981696 &  0.06121342 & -0.56934442 & -0.24230902 & -0.93795018 &  0.24806410 \\
Elqui           &  0.74099526 &  0.20483425 & -0.63950681 &  0.57756858 & -0.68021616 &  0.45135409 &  0.34255009 &  0.70381028 &  0.62234278 \\
Indus           &  0.47348784 & -0.22057954 & -0.85273321 &  0.25151201 & -0.89396596 &  0.37089969 &  0.84412734 &  0.39008914 &  0.36780360 \\
Jhelum          &  0.60334991 & -0.20211605 & -0.77143890 & -0.13408072 & -0.97928924 &  0.15170675 &  0.78612419 & -0.01190283 &  0.61795395 \\
Molonglo        &  0.88306113 &  0.15479520 & -0.44299152 &  0.36694639 & -0.81621072 &  0.44626270 &  0.29249510 &  0.55663139 &  0.77756550 \\
Phoenix         &  0.59644670 &  0.27151332 & -0.75533559 & -0.48595429 & -0.62682316 & -0.60904938 &  0.63882686 & -0.73032406 &  0.24192354 \\
Ravi            &  0.57336113 & -0.22475898 & -0.78787081 &  0.57203155 & -0.57862539 &  0.58135407 &  0.58654661 &  0.78401279 &  0.20319208 \\
Tucana III      &  0.505715   & -0.007435   & -0.862668   & -0.078639   & -0.996197   & -0.037514   &  0.859109   & -0.086811   &  0.504377   \\
Turbio          &  0.52548400 &  0.27871230 & -0.80385697 & -0.71193491 & -0.37328255 & -0.59481831 &  0.46584896 & -0.88486134 & -0.00227102 \\
Turranburra     &  0.36111266 & 0.85114984  & -0.38097455 &  0.87227667 & -0.16384562 &  0.46074725 & -0.32974393 &  0.49869687 &  0.80160487 \\
Wambelong       &  0.07420259 & 0.76149392  & -0.6439107  & -0.64686868 & -0.45466937 & -0.61223907 &  0.75898279 & -0.46195539 & -0.45884892 \\
Willka Yaku     &  0.37978305 & 0.29001265  & -0.87844038 & -0.5848418  & -0.66046543 & -0.47089859 &  0.71674605 & -0.69258795 &  0.08122206 \\
\enddata
{\footnotesize \tablecomments{ \matrixcomments }}
\label{tab:matrix}
\end{deluxetable}

\pagebreak

\bibliographystyle{aasjournal}
\bibliography{main}

\end{document}